\documentclass[twocolumn]{aastex62}
\usepackage{natbib}
\usepackage{color}
\usepackage{mathptmx}
\usepackage{amsmath}
\usepackage{url}
\usepackage{multirow}
\usepackage{verbatim}
\def\msun{\,{\rm M_\odot}}
\def\zsun{\,{Z_\odot}}
\usepackage{numprint}
\usepackage{graphicx}
\usepackage{refcount}
\graphicspath{{./}{figures/}}

\received{March 30, 2021}
\revised{\today}
\accepted{June XX, 2021}

\shorttitle{{\it AGORA} Comparison. III: Cosmological Zoom-in Simulation}
\shortauthors{{\it AGORA} Collaboration et al.}

\begin{document}
\title{The {\it AGORA} High-resolution Galaxy Simulations Comparison Project. III: Cosmological Zoom-in Simulation of \\ A Milky Way-mass Halo}

\author[0000-0002-6299-152X]{Santi Roca-F\`{a}brega}
\altaffiliation{Code leaders}
\correspondingauthor{Santi Roca-F\`{a}brega, sroca01@ucm.es}
\affil{Departamento de F\'{i}sica de la Tierra y Astrof\'{i}sica, Facultad de Ciencias F\'{i}sicas, Plaza Ciencias, 1, 28040 Madrid, Spain}
\author[0000-0003-4464-1160]{Ji-hoon Kim}
\altaffiliation{Code leaders}
\correspondingauthor{Ji-hoon Kim, mornkr@snu.ac.kr}
\affil{Center for Theoretical Physics, Department of Physics and Astronomy, Seoul National University, Seoul 08826, Korea}
\affil{Seoul National University Astronomy Research Center, Seoul 08826, Korea}
\author{Loic Hausammann}
\altaffiliation{Code leaders}
\correspondingauthor{Loic Hausammann, loic.hausammann@epfl.ch}
\affil{Institute of Physics, Laboratoire d'Astrophysique, \'{E}cole Polytechnique F\'{e}d\'{e}rale de Lausanne, CH-1015 Lausanne, Switzerland}
\author[0000-0001-7457-8487]{Kentaro Nagamine}
\altaffiliation{Code leaders}
\correspondingauthor{Kentaro Nagamine, kn@astro-osaka.jp}
\affil{Department of Earth and Space Science, Graduate School of Science, Osaka University, Toyonaka, Osaka, 560-0043, Japan}
\affil{Kavli IPMU (WPI), University of Tokyo, 5-1-5 Kashiwanoha, Kashiwa, Chiba, 277-8583, Japan}
\affil{Department of Physics \& Astronomy, University of Nevada Las Vegas, Las Vegas, NV 89154, USA}
\author{Alessandro Lupi}
\altaffiliation{Code leaders}
\affil{Dipartimento di Fisica ``G. Occhialini'', Universit\`a degli Studi di Milano-Bicocca, I-20126 Milano, Italy}
\author{Johnny W. Powell}
\altaffiliation{Code leaders}
\affil{Department of Physics, Reed College, Portland, OR 97202, USA}
\author{Ikkoh Shimizu}
\altaffiliation{Code leaders}
\affil{Shikoku Gakuin University, 3-2-1 Bunkyocho, Zentsuji, Kagawa, 765-8505, Japan}
\author{Daniel Ceverino}
\affil{Departamento de F\'{i}sica Te\'{o}rica, Facultad de Ciencias, Universidad Aut\'{o}noma de Madrid, Cantoblanco, 28049 Madrid, Spain}
\affil{CIAFF, Facultad de Ciencias, Universidad Aut\'{o}noma de Madrid, 28049 Madrid, Spain}
\author[0000-0001-5091-5098]{Joel R. Primack}
\affil{Department of Physics, University of California at Santa Cruz, Santa Cruz, CA 95064, USA}
\author{Thomas R. Quinn}
\affil{Department of Astronomy, University of Washington, Seattle, WA 98195, USA}
\author{Yves Revaz}
\affil{Institute of Physics, Laboratoire d'Astrophysique, \'{E}cole Polytechnique F\'{e}d\'{e}rale de Lausanne, CH-1015 Lausanne, Switzerland}
\author{H\'{e}ctor Vel\'{a}zquez}
\affil{Instituto de Astronom\'{i}a, Universidad Nacional Aut\'{o}noma de M\'{e}xico, A.P. 70-264, 04510, Mexico, D.F., Mexico}
\author{Tom Abel}
\affil{Kavli Institute for Particle Astrophysics and Cosmology, Stanford University, Stanford, CA 94305, USA}
\affil{Department of Physics, Stanford University, Stanford, CA 94305, USA}
\affil{SLAC National Accelerator Laboratory, Menlo Park, CA 94025, USA}
\author{Michael Buehlmann}
\affil{Laboratoire Lagrange, Universit\'{e} C\^{o}te d'Azur, Observatoire de la C\^{o}te d'Azur, CNRS, Blvd de l'Observatoire, CS 34229, 06304 Nice, France}
\author{Avishai Dekel}
\affil{Center for Astrophysics and Planetary Science, Racah Institute of Physics, The Hebrew University, Jerusalem 91904, Israel}
\author{Bili Dong}
\affil{Department of Physics, Center for Astrophysics and Space Sciences, University of California at San Diego, La Jolla, CA 92093, USA}
\author{Oliver Hahn}
\affil{Laboratoire Lagrange, Universit\'{e} C\^{o}te d'Azur, Observatoire de la C\^{o}te d'Azur, CNRS, Blvd de l'Observatoire, CS 34229, 06304 Nice, France}
\author{Cameron Hummels}
\affil{TAPIR, Department of Astronomy, California Institute of Technology, Pasadena, CA 91125, USA}
\author{Ki-won Kim}
\affil{Center for Theoretical Physics, Department of Physics and Astronomy, Seoul National University, Seoul 08826, Korea}
\author{Britton D. Smith}
\affil{Institute for Astronomy, University of Edinburgh, Royal Observatory, Edinburgh EH9 3HJ, United Kingdom}
\author{Clayton Strawn}
\affil{Department of Physics, University of California at Santa Cruz, Santa Cruz, CA 95064, USA}
\author{Romain Teyssier}
\affil{Centre for Theoretical Astrophysics and Cosmology, Institute for Computational Science, University of Zurich, Zurich, 8057, Switzerland}
\author{Matthew J. Turk}
\affil{School of Information Sciences and Department of Astronomy, University of Illinois, Urbana-Champaign, IL 61820, USA}
\author{the {\it AGORA} Collaboration}
\affil{\rm \url{http://www.AGORAsimulations.org}}
\affil{\rm The authors marked with * as code leaders contributed to the article by leading the effort within each code group to perform and analyze simulations.} 





\begin{abstract}

We present a suite of high-resolution cosmological zoom-in simulations to $z=4$ of a $10^{12}\,{\rm M}_{\odot}$ halo at $z=0$, obtained using seven contemporary astrophysical simulation codes ({\sc Art-I}, {\sc Enzo}, {\sc Ramses}, {\sc Changa}, {\sc Gadget-3}, {\sc Gear}, and {\sc Gizmo}) widely used in the numerical galaxy formation community. 
Physics prescriptions for gas cooling, heating and star formation are similar to the ones used in our previous {\it AGORA} disk comparison \citep{2016ApJ...833..202K_short} but now account for the effects of cosmological processes such as the expansion of the Universe, intergalactic gas inflow, and the cosmic ultraviolet background radiation emitted by massive stars and quasars. 
In this work, we introduce the most careful comparison yet of galaxy formation simulations run by different code groups, together with a series of four calibration steps each of which is designed to reduce the number of tunable simulation parameters adopted in the final run. 
In the first two steps, we methodically calibrate the gas physics such as cooling and heating, in simulations without star formation. 
In the third, we seek an agreement on the total stellar mass produced with the common star formation prescription used in the {\it AGORA} disk comparison, in stellar feedback-free simulations. 
In the last calibration step, we activate stellar feedback, where each code group is asked to set the feedback prescriptions to be as close to the most used one in each code community as possible, while aiming for convergence in the stellar mass at $z=4$ to the values predicted by semi-empirical models. 
After all the participating code groups successfully completed  the calibration steps, we reach a suite of cosmological simulations with similar mass assembly histories down to $z=4$.  
With numerical accuracy that resolves the internal structure of a target halo ($\lesssim 100$ physical pc at $z=4$), we find that the codes overall agree well with one another in e.g., gas and stellar properties, but also show differences in e.g., circumgalactic medium (CGM) properties. 
We argue that, if adequately tested in accordance with our proposed calibration steps and common parameters, the results of high-resolution cosmological zoom-in simulations can be robust and reproducible.
New code groups are invited to join and enrich this comparison by generating equivalent models or to test the code's compatibility on their own, by adopting the common initial conditions, the common easy-to-implement physics package, and the proposed calibration steps. 
Further analyses of the zoom-in simulations presented here will be in forthcoming reports from the {\it AGORA} Collaboration, including studies of the CGM, simulations by additional codes, and results at lower redshift. 

\end{abstract}

\keywords{cosmology: theory -- galaxies: formation -- galaxies: evolution -- galaxies: kinematics and dynamics -- galaxies: intergalactic medium -- galaxies: ISM -- methods: numerical -- hydrodynamics} 


\section{Introduction}\label{sec:intro}

\vspace{1mm}

Established in 2012, the {\it AGORA} High-resolution Galaxy Simulations Comparison Project ({\it Assembling Galaxies of Resolved Anatomy}) has since aimed at collectively raising the predictive power of contemporary numerical galaxy formation studies, by carefully comparing high-resolution galaxy simulations on multiple code platforms widely used in the field.
The main goal of the {\it AGORA} initiative has been to ensure that physical assumptions are responsible for any success in the numerical studies, rather than manifestations of a particular numerical implementation. 
As of this writing, we have more than 160 individuals from over 60 different academic institutions worldwide who have agreed to the Project's philosophy and participated in its collaborative effort in varying degrees. 
The Collaboration has continued to provide a sustainable platform on which members could talk to and learn from others from different code communities, and discuss ambitious ``multi-platform'' collaborations.  
The Project indeed has become a great social experiment in itself --- about the scientific community's collective willingness to assure the integrity and reproducibility of its experiments.\footnote{See the Project website at \url{http://www.AGORAsimulations.org/} for more information about the {\it AGORA} Collaboration.} 

The first paper of the Collaboration \citep[][hereafter Paper I]{2014ApJS..210...14K_short} focused on introducing the Project to the community.  
It presented the first proof-of-concept simulations, dark matter-only but using cosmological zoom-in initial conditions. 
Results from comparing the cosmological simulations among nine flavors of the state-of-the-art numerical codes showed a robust convergence.  
In the second paper from the {\it AGORA} Collaboration \citep[][hereafter Paper II]{2016ApJ...833..202K_short} we presented a comparison of idealized Milky Way-mass galaxies simulated in isolation, obtained from nine widely-used state-of-the-art gravito-hydrodynamics codes, which were recently made available to be freely used by the community \citep[][]{2020arXiv200104354R}. 
The simulations in Paper II achieved an overall agreement with one another in many parameter spaces for both gaseous and stellar components. 
Yet, some discrepancies were expected and present, which were understood as systematic differences between codes, for example, between mesh-based and particle-based codes in low-density regions, and between more diffusive and less diffusive schemes in the high-density region. 
Such intrinsic differences were, however, found to be small in general compared to the variations in the implementations of common subgrid physics such as supernova (SN) feedback.

The {\it AGORA} Project has helped to establish a simulation infrastructure essential to achieve our thorough comparisons so far, and it will allow and foster future comparisons. 
It includes, among others, a common initial condition generator \citep[{\sc Music};][]{2011MNRAS.415.2101H},\footnote{The website is \url{https://www-n.oca.eu/ohahn/MUSIC/}.} a common gas cooling and heating scheme \citep[{\sc Grackle};][]{2017MNRAS.466.2217S},\footnote{The website is \url{http://grackle.readthedocs.io/}.} and a common analysis toolkit \citep[{\tt yt};][]{2011ApJS..192....9T},\footnote{The website is \url{http://yt-project.org/}.} all of which are publicly available software.  
In particular, all the figures and plots  in this article and Papers I and II have been produced with the {\it AGORA} common analysis platform based on {\tt yt}.
It is also worth noting that several recent comparison and calibration studies have been motivated by the results presented in our previous reports. 
Examples include the study of changes on the star formation efficiency in molecular clouds \citep{2019MNRAS.486.5482G}, tests of new star formation and supernova feedback implementations, both in isolated \citep{Shimizu19} and cosmological contexts \citep{2020MNRAS.497.5203O}.

Building upon the past achievements, in this third paper of our continuing endeavor in {\it AGORA}, we follow a path similar to Paper II, but this time with cosmological ``zoom-in'' simulations. 
This type of comparison has never been properly carried out due to its complexity and time-consuming nature.  
However, it is now possible ---  though still challenging --- thanks to the infrastructure the {\it AGORA} Collaboration has built and maintained.
A reproducibility check like this is essential as the field relies increasingly on the numerical verification of galaxy formation theories in cosmological contexts.
All code groups started their simulations from a common initial condition generated with {\sc Music} (Section~\ref{sec:ic}). 
The physics prescriptions (e.g., gas cooling and heating, star formation parameters) are also common among all participating codes as in Paper II, although some changes were made in each code (Section~\ref{sec:physics}). 
Only the decision concerning the stellar feedback prescription and metal production to be used is left to each code group, and code groups are asked to use a prescription close to to the most widely-used practice in each code community. 
Spatial resolution of $\lesssim 100$ physical pc at $z=4$ is imposed  to resolve the internal structure of a target halo, and to make our physics prescriptions less reliant on platform-specific models (Section~\ref{sec:runtime}). 
After a series of calibration steps for the adopted physical processes  (Figure~\ref{FigFlowChart} and Section~\ref{sec:calibration}), we reach a suite of simulations illustrating how seven state-of-the-art codes reproduce the formation and evolution of a Milky Way-type galaxy in a cosmological context down to $z=4$ with their favorite stellar feedback and metal production prescriptions (Section~\ref{sec:results}). 
As in the previous {\it AGORA} comparisons, we caution that we do not intend to identify a correct or incorrect code, but to focus on juxtaposing different codes for physical insights and learn how much scatter one should expect among modern simulations.

This paper is organized as follows.
Section~\ref{sec:ic} describes the initial condition of our experiment. 
We discuss physics modules employed in our simulations in Section~\ref{sec:physics}, and the runtime parameters in Section~\ref{sec:runtime}. 
Section~\ref{sec:calibration} presents our calibration steps designed to prepare the ground for the final simulation entries.
In Section~\ref{sec:results} we compare the results of our final runs, focusing on the stellar and gas properties of the target halo, and its evolution in time.  
Finally, in Section \ref{sec:conclusion} we conclude the article with remarks on how {\it AGORA}'s ``multi-platform'' approach can significantly enhance the scientific value of numerical galaxy formation studies.

\vspace{1mm}

\section{Initial Condition} \label{sec:ic}

\vspace{1mm}

We use a set of parameters for {\sc Music}, an initial condition (IC)  generator with an adaptive multi-grid Poisson solver \citep{2011MNRAS.415.2101H}, that depicts a halo evolving to a virial mass of $\sim 10^{12} \,\,{\rm M}_{\odot}$ at $z=0$ with a relatively quiescent merger history between $z=2$ and 0.\footnote{Here we use {\sc Music}'s changeset ID {\tt eb870ed}.}
The IC, tagged {\it 1e12q}, is identified and made publicly available by the {\it AGORA} Collaboration (Paper I).\footnote{See \url{http://www.AGORAsimulations.org/} or \url{http://sites.google.com/site/santacruzcomparisonproject/blogs/quicklinks/}.}  
We assume a flat $\Lambda$CDM cosmology consistent with {\it WMAP7/9+SNe+BAO}: $\Omega_{\rm m}=0.272$, $\Omega_\Lambda=0.728$, $\sigma_8=0.807$, $n_{\rm s}=0.961$, and $H_0= 70.2\  {\rm km\ } {\rm s}^{-1} {\rm Mpc}^{-1}$ \citep{2011ApJS..192...18K_short, 2013ApJS..208...19H_short}.
The initial metallicity is set to $10^{-4} \zsun$ everywhere.\footnote{$1\, Z_{\odot} = 0.02041$ is used across all participating the codes in order to follow our choice in Paper II (see Section 2 of Paper II for details).}

With a $128^3$ root resolution in a $(60 \,\,\,{\rm comoving} \,\,h^{-1}\,{\rm Mpc})^3$ box and a series of five nested higher-resolution regions, the equivalent unigrid resolution at the finest ``zoom-in'' region is $4096^3$ (i.e., {\sc Music} parameters $\left[ \ell_{\rm min}, \ell_{\rm max} \right] = [7, 12]$).  
The highest-resolution region in this IC is in an ellipsoidal shape that is  large enough to enclose all the particles that eventually end up within $4 R_{\rm vir}$ of the target halo at $z=0$.
Correspondingly, the target halo contains the highest-resolution particles of masses ${m_{\rm DM, \,IC}} = 2.8\times10^5 \msun$ and ${m_{\rm gas, \,IC}} = 5.65\times10^4 \msun$ each, the latter designed to approximately match the gas resolution in Paper II, ${m_{\rm gas}} = 8.6\times10^4 \msun$. 
For more information about this IC and other available {\it AGORA} ICs, we refer the interested readers to Section 2 of Paper I.

\vspace{1mm}

\section{Physics In The Codes} \label{sec:physics}


We briefly summarize the key physics and code-by-code differences for this particular comparison.  


\subsection{Common, Code-independent Physics} \label{common-phy}


The common baryonic physics for our study is based on Papers I and II.  
To begin with, the cooling library {\sc Grackle} determines the rate of radiative gas cooling based on the properties of gas parcels \citep{2017MNRAS.466.2217S}.
The interface we built for Paper II is utilized by each participating code, in the {\it equilibrium} cooling mode of {\sc Grackle-v3.1.1}. 
Here, {\sc Grackle} looks up a pre-computed {\sc Cloudy} cooling table for primordial and $1 \zsun$ metallicities as  functions of gas density and temperature \citep{2013RMxAA..49..137F}. To obtain the corresponding gas cooling and heating rates, the $1 \zsun$ rates are linearly scaled by the gas metallicity (Section 3.1 of Paper II), and the result is added to the values from primordial gas to get the combined rate. 
{\sc Grackle} also includes redshift-dependent cosmic ultraviolet background radiation \cite[UVB;][]{HaardtMadau12} with hydrogen self-shielding (i.e., input file {\tt CloudyData\_UVB=HM2012\_shielded.h5}; see also Section 3.3 of Paper I). 
In  addition, instead of using {\sc Grackle}'s  own cosmic microwave  background  (CMB) temperature floor, each code  is  supplemented with a redshift-dependent, but density-independent CMB floor.\footnote{This functionality is planned to be added to the latest {\sc Grackle}.}  

Lastly, in order to prevent unphysical collapse or fragmentation due to limited resolution, in Calibration steps 3 and 4 we apply a nonthermal pressure floor $P_{\rm \,Jeans}$ that forces the local Jeans length to be resolved at a given numerical resolution at all times.  
Its value is $P_{\rm \,Jeans} = (\gamma \pi)^{-1} N_{\rm Jeans}^2 G \rho_{\rm gas}^2  \Delta x^2$,
where $\gamma=5/3$ is the adiabatic index, $N_{\rm Jeans} = 4$ is the Jeans number, $G$ is the gravitational constant, $\rho_{\rm gas}$ is the gas density, and $\Delta x$ is the finest spatial resolution in physical units (finest cell size for mesh-based codes, or gravitational softening length for particle-based codes; see Section \ref{sec:runtime}).  
This additional pressure term can be interpreted as the extra pressure source due to the unresolved interstellar medium (ISM) turbulence.
For actual implementations of the pressure floor in each code, we refer the readers to Sections 3.1 and Appendix A of Paper II.  

When the density of a gas parcel exceeds $n_{\rm H, \,thres} = 1\, {\rm cm}^{-3}$ (note the difference with $n_{\rm H, \,thres}$ used in Paper II), a star particle can be created  at a rate of
$d\rho_{\star}/dt ={\epsilon_\star \rho_{\rm gas} / t_{\rm ff}}$,
where $\epsilon_\star = 0.01$ is the formation efficiency and $t_{\rm ff}=(3\pi/(32 G\rho_{\rm gas}))^{1/2}$ is the local free-fall time. The only freedom that is left to each code group is to choose the stochastic or deterministic nature of this process. 
A single star particle depicts a collection of cluster-sized masses sharing the same age and metallicity, corresponding to a single stellar population.  
It is required to weigh more than $6.1\times10^4 \msun$ at creation for the mesh-based codes --- a value approximately matching the gas resolution in the IC, ${m_{\rm gas, \,IC}} = 5.65\times10^4 \msun$ --- or inherits the mass of its parent gas particle in particle-based codes.  
In Paper II, our stellar feedback formula implied one Type II supernova event per every 91 $\msun$ stellar mass formed, each of which instantaneously releasing $10^{51}$ ergs of thermal energy, 14.8 $\msun$ of gas, and 2.6 $\msun$ of metals.  
In contrast, in this work, while the returned mass is equal to that in Paper II, the exact deposit scheme into the ISM, such as stellar  winds or supernova events, and their associated energy and metal yields are left to each code group's discretion. 
We do ask the deposit scheme to be as close to the {\it most widely-used practice} in its  community as possible (detailed in Section \ref{code-phy} and Table \ref{tab:feedback}; see also Sections 3.2, 5 and Appendix B of Paper II).  
We also leave the choice of whether to implement an explicit metal diffusion scheme to each particle-based code group (see Sections \ref{sec:code-changa} to \ref{sec:code-gear} for details).  

We note that our common physics models including subgrid physics (e.g., star formation) helped us in Paper II to produce similar stellar disks across all codes --- comparable in terms of their morphologies, kinematics, star formation relations, to name a few (Sections 6.4 to 6.6 of Paper II). 
In the present comparison, however, we use a fully cosmological setup that is substantially more complex. 
Although the common subgrid physics models here are based on the ones in the idealized galaxy setup (Paper II), we have found a need to introduce changes to the fiducial parameters to reproduce a realistic galactic system at low redshift. 
The fiducial set of parameters has been modified in e.g., the star formation threshold density ($n_{\rm H, \,thres} = 1\, {\rm cm}^{-3}$ instead of $10\, {\rm cm}^{-3}$ in Paper II) and the stellar feedback scheme (instead of a common simple thermal deposit model in Paper II; see Section~\ref{code-phy}). 
These changes have been motivated by the deviation in $M_*/M_{\rm halo}$ from the observed value in Paper II, and to account for the potential redshift dependence of the adopted physics.

\vspace{1mm}

\subsection{Participating Codes and Code-dependent Physics} \label{code-phy}


Here we briefly explain the physics included in each code, focusing only on the part that is changed from Paper II, or is unique for each code.  
Hence, the interested readers are encouraged to see our previous work to grasp the full picture of how each code works --- Paper I for gravitational dynamics and Paper II for hydrodynamics.   
In particular, Table~\ref{tab:feedback} summarizes the key stellar feedback parameters and effective metal yield in each code, in which each code group is left with freedom to choose its own feedback scheme for energy and metals.
It should be noted that the code groups involved in future {\it AGORA} studies are not limited to the seven codes listed in this section.

\begin{table*}
\vspace*{1mm}
\footnotesize
\caption{\footnotesize Stellar feedback implementation adopted by each code group\tablenotemark{\textdagger}}
\centering
\begin{tabular}{c || c | c | c | c }
\hline\hline 
Code & Stellar feedback & SN \& metal production model & Effective metal yield & Runtime parameters\\ 
\hline
{\sc Art-I} & T+K, RP  & SN Type Ia/II, AGB stars$^*$ & 0.033 & $E_{\rm thermal}= 2\times 10^{51}\,{\rm ergs/SN}$,\, $p =3.6 \times 10^6 \msun \, {\rm km \, s^{-1}/SN}$  \\
{\sc Enzo} & T  & SN Type II  & 0.032 & $E_{\rm thermal} = 5\times 10^{52}\,{\rm ergs/SN}$ \\ 
{\sc Ramses} & T, DC  & SN Type II & 0.033 & $E_{\rm thermal}= 4\times 10^{51}\,{\rm ergs/SN}$, $\sigma_{\rm min}=100\,\,\rm{km \, s^{-1}}$, $\,\,T_{\rm delay}=10\,\,\rm{Myr}$  \\
{\sc Changa} & T+S & SN Type Ia/II, AGB stars$^{**}$ & 0.032 & $E_{\rm thermal} = 5\times10^{51}\,{\rm ergs/SN}$  \\  
{\sc Gadget-3} & T+K, RP, DC  & SN Type Ia/II, AGB stars & 0.025 & $E_{\rm SN} = 4\times 10^{49}\,{\rm ergs/ \msun}, \,\,\,T_{\rm delay}=t_{\,\rm hot}$ (see Section \ref{sec:code-gadget})  \\  
{\sc Gear} & T, DC  & SN Type Ia/II & 0.024 & $E_{\rm thermal} = 4.5\times10^{51}\,{\rm ergs/SN}$, $\,\,T_{\rm delay}=5\,\,\rm{Myr}$  \\
{\sc Gizmo} & T+K & SN Type II & 0.033 & $E_{\rm SN} = 5\times10^{51}\,{\rm ergs/SN}$  \\  
\hline 
\end{tabular}
\tablenotetext{$\textdagger$}{\scriptsize T = thermal feedback, K = kinetic feedback, RP = radiation pressure, DC = delayed cooling, S = superbubble, $^*$ = only for energy production (not metal), $^{**}$ = only for metal production (not energy).  While the total returned mass via feedback is constrained across the code platforms (Section \ref{common-phy}), the exact feedback scheme and the metal yield are left to each code group's discretion to be as close to the most widely-used practice in its community as possible.  For more information on the items listed here, see Section~\ref{code-phy}.  For more information on the ``effective'' metal yield by stellar feedback measured in the entire simulation box at $z=4$ for the {\tt CosmoRun} suite of simulations (fourth column), see Section~\ref{gas-metallicity}.}
\label{tab:feedback}
\vspace*{2mm}
\end{table*}

\subsubsection{\sc Art-I} \label{sec:code-art}

The {\sc Art-I} code \citep{1997ApJS..111...73K,2003ApJ...590L...1K,2009ApJ...695..292C} used to obtain the cosmological simulation presented here is based on the one used in the previous comparison efforts (Papers I and II). 
Only a few minor modifications should be noted.  
Among them is a change in the adaptive mesh refinement (AMR) strategy to better follow the cosmic evolution of large scale structures. 
This change is in line with what has been commonly used in previous {\sc Art-I} cosmological zoom-in simulations \citep[e.g.,][]{2010MNRAS.404.2151C, 2014MNRAS.442.1545C, 2017MNRAS.467.2664C}.  
We have also updated the gas cooling and heating scheme from {\sc Art-I}'s own machinery using the {\sc Cloudy} table in Paper II to the standard package {\sc Grackle-v3.1.1} in the current paper. 
The nonthermal pressure floor in {\sc Art-I} is slightly different from the common prescription (Section \ref{common-phy}); in other words, the Jeans length is resolved by at least seven resolution elements at all times \citep{2010MNRAS.404.2151C}.  

{\sc Art-I} uses a stochastic star formation subgrid model.  
Details on this star formation model can be found in \citet{2014MNRAS.442.1545C}. 
We slightly change the stochasticity of star formation to ensure that we use the common star formation efficiency value (Section \ref{common-phy}).
{\sc Art-I}'s prescription fits within the agreed {\it AGORA} parameter range.
The treatment of stellar feedback is similar to the model in \citet{2017MNRAS.467.2664C}, which includes thermal,  kinetic and radiation pressure feedback.
The code also includes the later effects of supernova Type Ia and stellar mass loss, and it follows the metal enrichment of the ISM.
The convergence goal in the calibration step 4 ({\tt Cal-4}; Section \ref{sec:cal_4}) is achieved by the widely used feedback model in the {\tt VELA6} simulations (Ceverino et al, in prep.)
but with four times more injection of momentum (see parameter $p$ in Table \ref{tab:feedback}). This increase tries to compensate for the differences in resolution. The default AGORA effective metal yield has been obtained by increasing the standard SNII and SNIa yields in ART-I by a factor of four.

\subsubsection{\sc Enzo} 

The {\sc Enzo} code  \citep{2014ApJS..211...19B, Brummel-Smith2019} for this work is  from the {\tt master} branch in the publicly available {\tt enzo-dev} repository.\footnote{The website is \url{http://enzo-project.org/}.  Here we use {\sc Enzo}'s changeset ID {\tt 02c88172}.} Star formation is implemented following the same approach as in Paper II, that is a fully deterministic scheme.
To incorporate the stellar feedback model established in Paper II, files such as  {\tt star\_maker4.F} and {\tt Grid\_StarParticleHandler.C} in the said repository need a minor modification.  
To reach the convergence in our calibration step 4 ({\tt Cal-4}; Section \ref{sec:cal_4}), the stellar feedback efficiency parameter is increased from the value in Paper II, matching the findings in recent {\sc Enzo} calibration studies against observations \citep[e.g.,][see also Table~\ref{tab:feedback}]{2020MNRAS.497.5203O}. 
The model only accounts for effects by supernova Type II.
Other adopted schemes such as the hydrodynamics solver are the same as in Paper II, and are largely in line with the recent numerical galaxy formation studies using {\sc Enzo} \citep[e.g.,][]{2019ApJ...887..120K, 2020ApJ...899...25S}. 

In order to realize the ellipsoid-shaped IC in simulations (Section~\ref{sec:ic}), {\sc Enzo} identifies and tracks the ellipsoidal Lagrangian region using a special type of dark matter particles called {\tt MustRefineParticle} that eventually constitute the target halo at a predetermined target redshift.
Cells around these particles are always refined at least down to 20.9 comoving kpc --- or 5 additional refinement levels for a $128^3$ root resolution in a $(60 \,\,\,{\rm comoving} \,\,h^{-1}\,{\rm Mpc})^3$ box --- corresponding to the {\sc Music} parameter $\ell_{\rm max} = 12$.  

\subsubsection{\sc Ramses}

The {\sc Ramses} code \citep{Teyssier2002} used in this comparison is from the December 2019 {\tt master} branch of the code repository.\footnote{The website is \url{https://bitbucket.org/rteyssie/ramses/}.} 
Star formation is implemented following Paper II, but without using a temperature threshold. 
This temperature threshold was closely linked with the implementation of a temperature polytrope to avoid numerical fragmentation, and this approach is no longer in use in the present work.
Thus, the implementation of the nonthermal pressure support to avoid artificial fragmentation takes a different approach from the one in Paper II, being now consistent with the common implementation presented in Section~\ref{common-phy}. With this implementation we ensure that the local Jeans length is resolved at least by four AMR cells at all times.
The star formation approach is well described in the most recent works within the code community \citep[e.g.,][]{2020arXiv200406008N}. 

The treatment of stellar feedback here closely follows the so-called ``delayed cooling thermal feedback model''  formulated in \citet{2015MNRAS.452.1502D}, and only accounts for effects by supernova Type II.
The {\sc Ramses} simulation presented here includes modifications to the model, however, as described in  \citet[][Section 3.3]{2017MNRAS.466...11R} and \citet[][Section 2.1.3]{2020arXiv200406008N}.  
Our choices of runtime parameters are listed in the Table~\ref{tab:feedback}.
We note that, out of our tested feedback prescriptions available in {\sc Ramses}, the one used here is what succeeded in producing the target stellar mass at $z=4$ in our calibration step 4 ({\tt Cal-4};  Section \ref{sec:cal_4}). 

\subsubsection{\sc Changa} \label{sec:code-changa}

{\sc Changa-v3.4} is a reimplementation of the smoothed particle hydrodynamics (SPH) code {\sc Gasoline} \citep{Wadsley2017} in the {\sc Charm++} {\sc (CharmppPPWCPP96)} v6.9 runtime system.\footnote{The websites are \url{http://github.com/N-BodyShop/changa/} and \url{http://charm.cs.uiuc.edu/}.} 
The code used for the present paper is based on the one in the previous hydrodynamic comparison; therefore, we refer the interested readers to Section 5.5 of Paper II and here we note only a few points and changes.
In {\sc Changa}, the $k$-th nearest neighbor algorithm is used to find the $N_{\rm ngb} = $ 64 nearest neighbors, then the Wendland C4 kernel \citep{Dehnen2012} is employed to determine hydrodynamic properties. 
Energy and metals are diffused using the scheme of \citet{Shen2010}.  
We have implemented {\sc Grackle-v3.1.1} after careful scrutiny.\footnote{The {\sc Cloudy} table used in {\sc Changa} differs slightly from the one in the other codes, containing a latest update by the {\sc Grackle} developers.  This update only affects an unlikely case of very dense gas at very high redshifts, so it does not change the conclusion of the present article.}

The treatment of stellar feedback follows the ``superbubble'' strategy presented by \citet{Keller2014}, different from Paper II. 
It includes thermal conduction inside resolved hot bubbles, which maintains uniform temperatures (see the characteristic bubble shapes in Figure \ref{Fig14}).
This method makes the amount of cold gas heated by feedback not a free parameter, but set by the thermal conduction. 
In the first few Myr of feedback heating, the mass contained within a hot bubble can be smaller than the simulation's gas mass resolution, which could result in strong overcooling. 
To prevent overcooling, the resolution elements briefly represent two components: (1) a hot interior (bubble) where the feedback energy is injected, and (2) a cold shell in pressure equilibrium with the hot interior. 
The particle returns to a single phase once all the cold gas is evaporated or the hot phase cools below $10^5$ K. 
Thermal energy representing supernova Type Ia and Type II is deposited to the neighboring $N_{\rm ngb}$ particles. 
Supernova Type II rates are calculated from the \citet{Raiteri1996} fit to the Padova stellar models. 
Type Ia rates are computed from the evolution timescales of secondaries in binaries \citep{Matteucci1986}. 
To reach the convergence in our calibration step 4 ({\tt Cal-4}; Section \ref{sec:cal_4}), the thermal energy is increased to $5\times 10^{51}$ ergs per supernova for the Kroupa  initial mass function (IMF), from the typical value used in the community.
Metals are released by supernovae and asymptotic giant branch (AGB) stars following \citet{Raiteri1996}. 

\subsubsection{\sc Gadget-3} \label{sec:code-gadget}

{\sc Gadget-3-Osaka} is a modified version of {\sc Gadget-3} --- which itself is an extended version of the SPH code {\sc Gadget-2} \citep{2005MNRAS.364.1105S}.  
The code includes the common cooling and star formation model detailed in Papers I and II, and the treatment of stellar feedback presented in \cite{2017MNRAS.466..105A, 2018MNRAS.478.4905A} and \cite{Shimizu19}.  
It also includes important improvements such as the density-independent, pressure-entropy formulation of SPH \citep[][]{hopkins_general_2013,Saitoh2013}, the time-step limiter \citep{Saitoh09}, 
quintic spline kernel \citep{Morris96}, and the number of neighbor particles for each SPH particle is set to $128\pm 8$.

For stellar feedback, we distribute both thermal and kinetic energy to neighboring gas particles within a hot bubble, whose size is determined by the local gas density, ambient gas pressure, and feedback energy (see Eqs.(6)-(7) in \citealt{Shimizu19}). 
We utilize the {\sc CELib} chemical evolution library \citep[][]{2017AJ....153...85S} which provides the chemical yield distribution as a function of time for a given IMF.    
We deposit metals and  energy according to the {\sc CELib} output with certain time delays ($t_{\,\rm hot}$) that depend on the feedback energy, density, and ambient gas pressure, treating supernova Type Ia, Type II, and AGB star contributions separately.\footnote{For example, oxygen production is always dominated by Type II supernova, carbon is dominated by AGB stars after a few hundred Myrs, and iron is dominated by Type Ia supernova after $10^8$ years.} 
The total injected energy is slightly boosted over the canonical {\sc CELib} output, to $4\times 10^{49}\,{\rm ergs}$ per 1\,$\msun$ of star forming gas, corresponding to $E_{\rm SN}= 4\times 10^{51}$\,erg per supernova for the Chabrier IMF adopted in {\sc CELib}.
For details, see \cite{Shimizu19}. 
The exact prescription used in this paper is similar to the fiducial model {\tt K30T70} therein, except for the equal division of  supernova energy into thermal (50\%) and kinetic (50\%) component to reach the target stellar mass ({\tt Cal-4}; Section \ref{sec:cal_4}).  
Early stellar feedback is also adopted in the form of thermal energy injected before the first supernova explodes. 
Metal diffusion is not implemented as an explicit process, but metals are smoothed over the SPH kernel when computing the metallicity or cooling rates of each gas particle, mimicking the effect of metal diffusion \citep{okamoto_effects_2005, tornatore_chemical_2007, wiersma_chemical_2009}. 

\subsubsection{\sc Gear} \label{sec:code-gear}

The {\sc Gear} code is a chemo-dynamical tree SPH code based on {\sc Gadget-2} \citep{2005MNRAS.364.1105S}. 
Its original version was described in \citet{revaz_dynamical_2012} with some improvements discussed in \citet{revaz_computational_2016} and \citet{revaz_pushing_2018}. 
For the difference between {\sc Gear} and  the public version of {\sc Gadget-2}, we refer the interested readers to Section 5.8 of Paper II.  
Cooling and star formation prescriptions adopted here are similar to the ones in Paper II.

In our feedback prescription, both energy and yields are deposited among the nearest gas particles so that each neighbor receives a fraction of energy weighted by the SPH kernel.  
$N_{\rm ngb}$ corresponds to a {\it weighted} number of neighbors and is set to 50. 
Thus, depending on the spatial distribution of gas particles more or less than 50 particles will receive stellar ejecta.
The stellar feedback is tightly coupled to our adopted chemical evolution model, that includes both supernova Type Ia and II with yields from \citet{kobayashi_history_2000} and  \citet{tsujimoto_relative_1995}, respectively.
Exploding supernovae are computed stochastically using a continuous IMF sampling scheme \citep[CIMFS;][]{revaz_computational_2016}.
Thus here, a thermal energy equivalent to $4.5 \times 10^{51} \,{\rm ergs}$ per supernova  is released into the ISM,  following a blast wave-like feedback scheme \citep{2006MNRAS.373.1074S} with a 5 Myr delayed cooling time. 
While {\sc Gear} does not include artificial metal diffusion, we use the smooth metallicity scheme to mix the metal-enriched gas effectively (as in {\sc Gadget-3}; see Sections \ref{sec:code-gadget}).

\subsubsection{\sc Gizmo} \label{sec:code-gizmo}
\textsc{Gizmo} is a mesh-free hydrodynamics code \citep{2015MNRAS.450...53H},  a descendant of \textsc{Gadget-3}, in which a kernel-based partition scheme is used to discretize the domain in a set of unstructured ``cells'' that are allowed to move and reshape with time. 
The Riemann problem is solved across the effective faces shared by neighbouring cells, similarly to what is done in the grid-based codes. 
The version used for this work includes the common cooling and star formation models described in Paper II while stellar feedback is based on the mechanical feedback model described in \citet{2018MNRAS.477.1578H}; i.e., both kinetic and thermal energy are distributed among gas cells lying within each star particle kernel according to the evolutionary stage of the supernova blast-wave (energy or momentum conserving). 
The supernova rate used in this work is described by a piecewise function, where we assume the decaying power-law fit in \citet{Lupi_2020} for star particles older than 5.089~Myr, and a constant rate equal to the power-law maximum value for younger stars, aimed at modelling early feedback by massive stars. 
For consistency, the integrated number of supernova events is normalised to ensure 1 supernova per every 91$~\rm M_\odot$, while the injected energy is set to $5\times 10^{51}$~ergs per supernova in order to reproduce the desired stellar mass at $z=4$. 
\vspace{1mm}

\section{Common Runtime Parameters} \label{sec:runtime}


We describe our choices of common runtime parameters such as numerical resolution.  
They are based on what we used in the dark matter-only cosmological test for a galaxy-sized halo (Section 5 of Paper I), and in the isolated disk test in a Milky Way-sized halo (Section 4 of Paper II).  

\begin{figure}
        \centering
        \vspace{1mm}
        \includegraphics[scale=0.58]{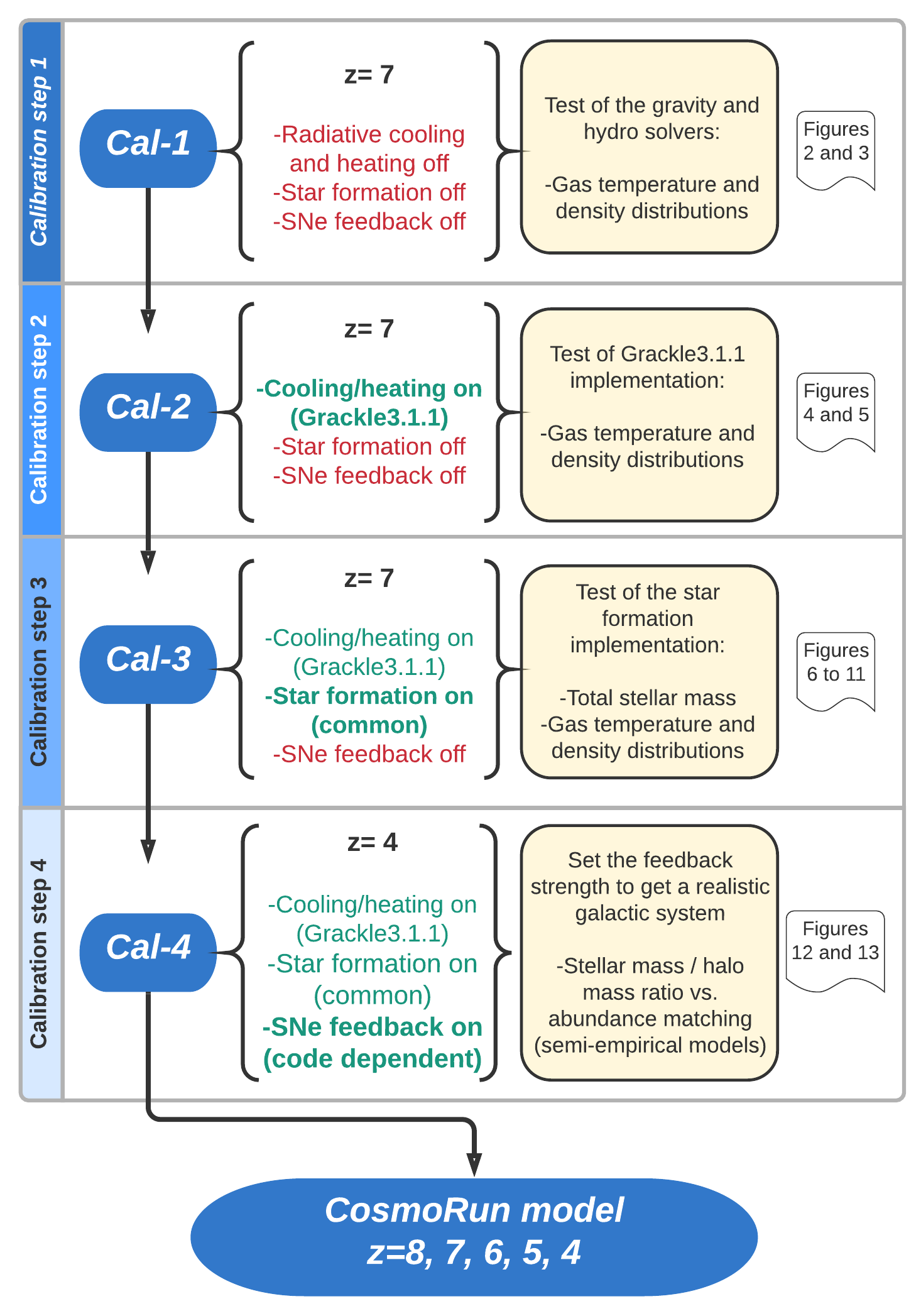}
        \caption{Summary of the physics calibration procedure. 
        We indicate, from left to the right, the target redshift and the physics prescriptions in each step, the main objective and the used variables to test convergence, and the corresponding figures.}
        \label{FigFlowChart}
        \vspace{1mm}
\end{figure}

\begin{figure*}
        \centering
        \vspace{0mm}
        \includegraphics[scale=0.6]{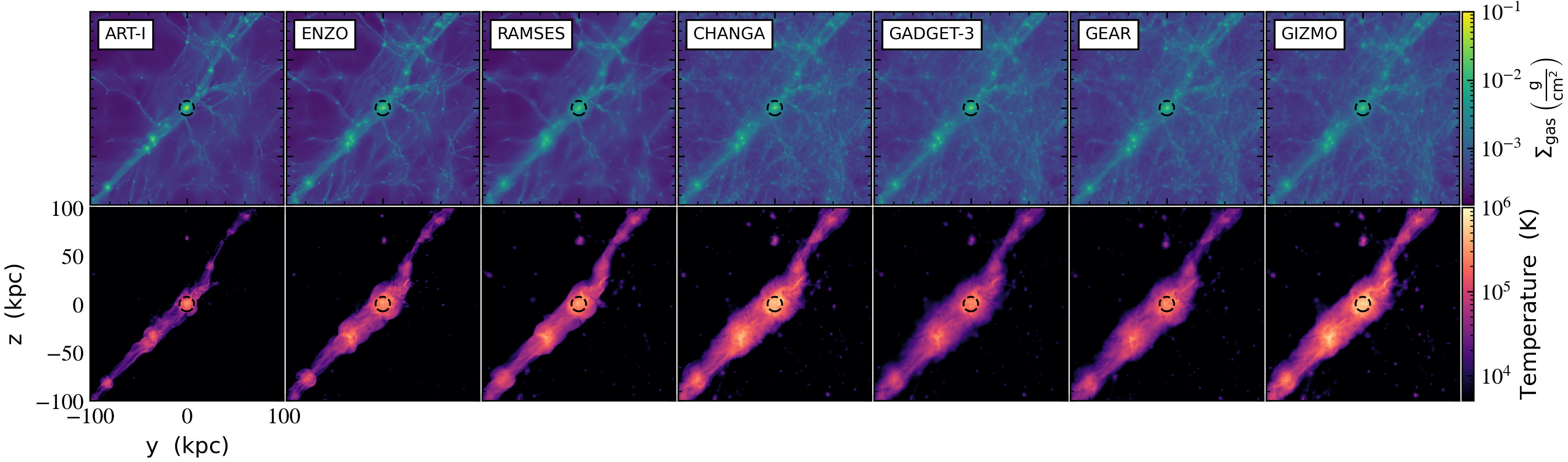}  
        \vspace{0mm}      
        \caption{Gas density projection ({\it top}) and density-weighted temperature projection ({\it bottom}; each projected through a slab of thickness 200 kpc) at $z=7$ from the first calibration step, {\tt Cal-1} (adiabatic evolution test). 
        We indicate the mean $R_{200}$ among the codes ($\sim 7.5$ kpc at $z=7$) with a black dashed circle. 
        Units are proper kpc. 
        See Section \ref{sec:cal_1} for more information on {\tt Cal-1} and this figure, and Section \ref{code-phy} for descriptions of participating codes in this comparison.  
	The full color version of this figure is available in the electronic edition.  
	The high-resolution versions of this figure and article are available at the Project website, \url{http://www.AGORAsimulations.org/}.
	Simulations are performed by:  Santi Roca-F\`{a}brega ({\sc Art-I}, {\sc Ramses}), Ji-hoon Kim ({\sc Enzo}), Johnny Powell and H\'ector Vel\'azquez ({\sc Changa}), Kentaro Nagamine and Ikkoh Shimizu ({\sc Gadget-3}), Loic Hausammann and Yves Revaz ({\sc Gear}), Alessandro Lupi and Bili Dong ({\sc Gizmo}).  
	}
        \label{Fig00}
        \vspace{5mm}
\end{figure*}

For the particle-based codes {\sc Changa}, {\sc Gadget-3} and {\sc Gear}, a spline kernel  is used to soften the gravity \citep[e.g., Eq. (A1) of][]{1989ApJS...70..419H}.  
The gravitational softening length $\epsilon_{\rm grav}$ in the highest-resolution region is set to 800 comoving pc until $z=9$, and 80 proper pc afterward. 
While this resolution is better than what Eq. (15) of \cite{2003MNRAS.338...14P} proposes ($\sim 220$ pc), it is used to match the resolution of Paper II at which our fiducial subgrid physics models were initially calibrated.   
For particles in the lower-resolution region at a corresponding {\sc Music} level $\ell$, the softening length is set at $80 \times 8^{(\ell_{\rm max}-\ell)/2}$ proper pc after $z=9$, as \cite{2003MNRAS.338...14P} suggests $\epsilon_{\rm grav,\, \ell} \propto N_{200}^{\,-1/2} \propto (m_{\rm DM, \,\ell})^{1/2}$.
For particle-based codes, we also require that the minimum hydrodynamical smoothing lengths for gas particles be $0.2\,\, \epsilon_{\rm grav}$.
The exact choice for a smoothing scheme is left to each code group's discretion (see Section 5 and Appendix C of Paper II). 

Meanwhile, the finest cell size of the mesh-based codes  {\sc Art-I}, {\sc Enzo} and {\sc Ramses}) is set to 163 comoving pc, or 12 additional refinement levels for a $128^3$ root resolution in a $(60 \,\,\,{\rm comoving} \,\,h^{-1}\,{\rm Mpc})^3$ box.   
A cell is adaptively refined into 8 child cells on particle or gas over-densities of 4.  
Given the differences in refinement algorithms among the codes, parameters that control the overall mesh structure and the aggressiveness of the refinement are left for each code group to decide (see Section 5 of Paper II). 
These differences can have an impact on the gas density and temperature distributions when without stellar feedback (as shown in Sections~\ref{sec:cal_1} and \ref{sec:cal_2}), but the impact becomes marginal once stellar feedback is activated (Section~\ref{sec:cal_4}). 
Further analyses of such differences in the evolution of primordial gas at high $z$ will be presented in future papers from the {\it AGORA} Collaboration.

Lastly, we recommend that each group stores simulation outputs at 200 epochs.\footnote{200 epochs starting from $a=0.062\,\, (z\sim15)$ to $a=0.325\,\, (z\sim2)$, equally spaced  in log(a) with $\Delta {\rm log(a)} = | {\rm log}(384/2013)/200 |$, plus a set of redshift snapshots at $z = 15, \,14, \,13, \,12, \,11, \,10, \,9, \,8, \,7, \,6, \,5, \,4, \,3, \,2$.  
Downloadable  at \url{http://physics.snu.ac.kr/cosmo/agora/output\_z\_cosmorun.txt}.}
An explicit list of this {\it AGORA}-recommended output interval is publicly available, and can be used by anyone to compare their simulation with {\it AGORA}.

\vspace{1mm}

\section{Physics Calibration Steps}  \label{sec:calibration}


Before proceeding to generate the final cosmological simulations, all participating code groups have been asked to complete four rigorous calibration steps. 
The main objective of these calibrations is to reduce the number of free parameters and artifacts in each code that can have an impact on the evolution of simulated galaxies, that are not valid physical assumptions about the structure formation. 
By adding {\it one physical process at a time} into our cosmological zoom-in simulation, we seek a situation where all code groups converge to a final simulation with similar global properties (e.g., similar stellar mass) --- and thus, any differences can only be attributed to the chosen stellar feedback prescriptions and intrinsic variations of the codes' numerics. 
We summarize the calibration procedure with a flowchart in Figure~\ref{FigFlowChart}.

The first two calibration steps (hereafter {\tt Cal-1} and {\tt Cal-2}) are designed to first acquire qualitative convergence on the main gas properties, by calibrating the gas physics such as cooling and heating when star formation is not enabled. 
In the third calibration step ({\tt Cal-3}), with star formation enabled, but the corresponding stellar feedback disabled, we look for agreement in the main gas properties and in the total stellar mass produced at $z=7$. 
Finally, in the fourth step ({\tt Cal-4}), we activate stellar feedback and aim to achieve convergence only in the stellar mass at $z = 4$ to the values predicted by semi-empirical models.  
Each code group is asked to set the feedback prescriptions to be as close to the most used one in each code community as possible.  
This last calibration step is a groundwork from which we can study how galactic properties  depend on feedback prescriptions.

An important result of our set of calibrations is that the simulation parameters selected in an isolated disk test (Paper II) cannot be naively used in the cosmological simulations like the ones presented here. 
Gas properties (e.g., metallicity) and the external radiation field rapidly evolve with redshift, which has a strong impact on gas cooling and, thus, star formation. 
Furthermore, continuous acquisition of fresh gas from the intergalactic medium (IGM) and circumgalactic medium (CGM) makes the cosmological run substantially more complex than that of an isolated disk galaxy.

In this section, we carefully describe the four calibration steps one by one.  
We start each subsection by explaining its setup, and then go through the important findings and conclusions from each step.  
One could consider each of our calibration steps as a standalone comparison in itself.  
Nevertheless, when successively executed and combined with other steps, our calibration procedure provides a solid ground on which advanced cosmological simulations could be performed and trusted.  
For example, new code groups may test their code's compatibility with the other contemporary codes, by following the common initial conditions, the common physics package, and the calibration steps proposed  herein.


\subsection{Calibration Step One \,({\tt Cal-1}): Adiabatic Evolution of Gas}\label{sec:cal_1}

The first calibration step we undertake ({\tt Cal-1}) is designed to detect inter-platform variations in the temperature and density of the accreted gas at $z=7$ when no radiative process or subgrid physics is present. 
Each cosmological run has been performed without any radiative cooling processes or heating sources, or any subgrid models such as star formation or the pressure floor.  
Under such conditions, the system exchanges no energy with its surroundings, and is considered adiabatic. 
The system's entropy, however, is not necessarily constant as it may increase owing to the presence of shocks.
If so, any variation between the codes is in principle caused only by the differences in hydrodynamics solvers --- namely, how each code solves the conservation laws of fluid dynamics and how shocks, e.g., in the accreting gas, are captured and treated. Despite small differences described below, an overall convergence has been found among the seven participating simulation codes.


\subsubsection{Findings From {\tt Cal-1}}\label{sec:Cal1_comp}

\begin{figure}
        \centering
        \vspace{2mm}
        \includegraphics[scale=0.25]{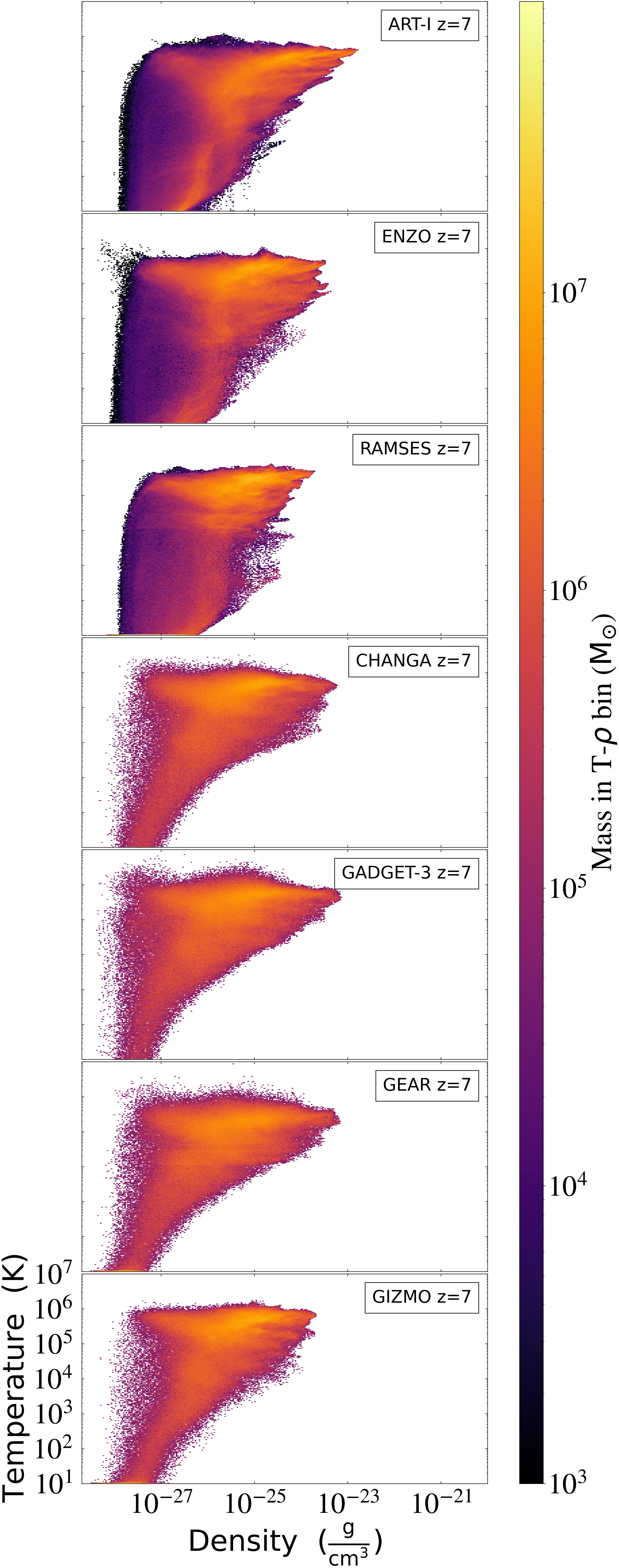}        
        \caption{The $z=7$ composite of 2-dimensional probability distribution function (PDF) of density and temperature for the gas within 100 kpc from the center of the main galactic system in the {\tt Cal-1} runs.  
        The 100 kpc-radius sphere encloses the main galaxy, the CGM, and the nearby IGM.  
        Colors represent the total gas mass in each 2-dimensional bin.  
        In all analyses for particle-based codes hereafter --- except the graphical visualizations such as Figures \ref{Fig00} or \ref{Fig5.0} --- raw particle fields are used, not the smoothed fields built by {\tt yt}.
        See Section \ref{sec:cal_1} for more information on {\tt Cal-1} and this figure.  
        }
        \label{Fig01}
\end{figure}

In Figure~\ref{Fig00} we show the projected density (top row) and temperature (bottom row) from {\tt Cal-1} at $z=7$. 
The virial radius, defined as $R_{200}$, is approximately 7.5 kpc at $z=7$ across all the codes  (see Table~\ref{tab:Generalparams}), shown as black dashed circles.
In Figure~\ref{Fig00} and similar projection images hereafter, particle-based codes are smoothed using a spline kernel in {\tt yt}.\footnote{We employ {\tt yt-v4.0} which better handles the SPH particles, an improvement from {\tt yt-v3.3} used in Paper II. 
See how  {\tt yt-v4.0}'s handling of SPH particles differs from that of its predecessors at \url{https://matthewturk.github.io/yt4-gallery/}.} However, these codes are not smoothed in other types of figures and analysis in this paper.
Meanwhile, in Figure~\ref{Fig01}, we show the density-temperature probability distribution function (PDF) at the same epoch for all the gas within 100 kpc from the center of the main progenitor. 
Because the virial radius of the target halo at this redshift is $\sim7.5$ kpc, we are showing a volume that includes gas not only in the galaxy, but also inside filaments, sheets, knots, and voids. 

Overall, the large-scale density structures in all seven panels of Figure~\ref{Fig00} are remarkably similar with one another,  
and multiphase density-temperature structures in Figure~\ref{Fig01} are also comparable.  
Unsurprisingly, in both plots, the convergence is very good qualitatively for the particle-based codes, {\sc Changa}, {\sc Gadget-3}, {\sc Gear}, and {\sc Gizmo} as they share gravity solvers and take similar SPH approaches. 
The three mesh-based codes, {\sc Art-I}, {\sc Enzo}, and {\sc Ramses}, show minor differences but an overall agreement, too. 
Larger discrepancies are observed when comparing particle-based codes with mesh-based codes.
In particular, the differences in the resolved structures in low-density regions at high redshift were discussed in the previous {\it AGORA} comparison with dark matter-only simulations.   
It is because particle-based codes achieve better resolution at early times than the mesh-based codes assuming little or no adaptive refinement for the mesh-based codes at high $z$ (for detailed discussion, see Section 5.3.2 of Paper I).  
We also notice in Figure~\ref{Fig01} that the highest densities that each code reaches are somewhat different, particularly among the mesh-based codes. 
This is due to differences in the refinement strategies adopted in each code, and we plan to study this issue further in future publications. 

\subsubsection{Comments On The Differences In The Warm-Hot Intergalactic Medium In {\tt Cal-1}}\label{Cal1_whim}

From Figure~\ref{Fig00}, one however notices some discrepancies in the temperature maps.
While all codes reproduce the virialized hot gas expected around massive haloes, with temperature between $10^5-10^6\,\rm{K}$, it is clear from 
Figure~\ref{Fig01} that the extension of this hot component to lower densities --- the warm gas that surrounds the main galactic systems --- slightly differs. 
In particular, in {\sc Art-I}, the intergalactic warm gas extends only up to the virial radius indicated in Figure~\ref{Fig00} by the dashed black circles, while it extends beyond the virial radius and encompasses more mass in {\sc Changa}, {\sc Gadget-3}, {\sc Gear}, and {\sc Gizmo}. 

The effects that accretion shocks have over the warm gas around the main galactic systems could be different between codes, as they can be caused by small differences in numerical techniques.  
This phenomenon has been documented by many authors: 
{\it (1)} Gas could be overheated via collisional heating with dark matter particles due to differences in gravity solvers, integrators, timestepping strategies for force calculations, and refinement strategies \citep{2010MNRAS.401..791S, Luki__2014, Jia_2020}. 
{\it (2)} Gas could be overheated also by the artificial viscosity in the sharp accretion shocks in particle-based codes \citep{2012MNRAS.423.1726S,2012MNRAS.426.1687T,2016ApJS..224...32H}. 
{\it (3)} Gas could be overcooled in the accretion shocks due to low resolution in the insufficiently refined CGM \citep{2013MNRAS.432..711H}. Although here we present the first analysis, this will be better characterized in a future paper from the Collaboration.

\begin{figure}
        \centering
        \vspace{2mm}
        \includegraphics[scale=0.25]{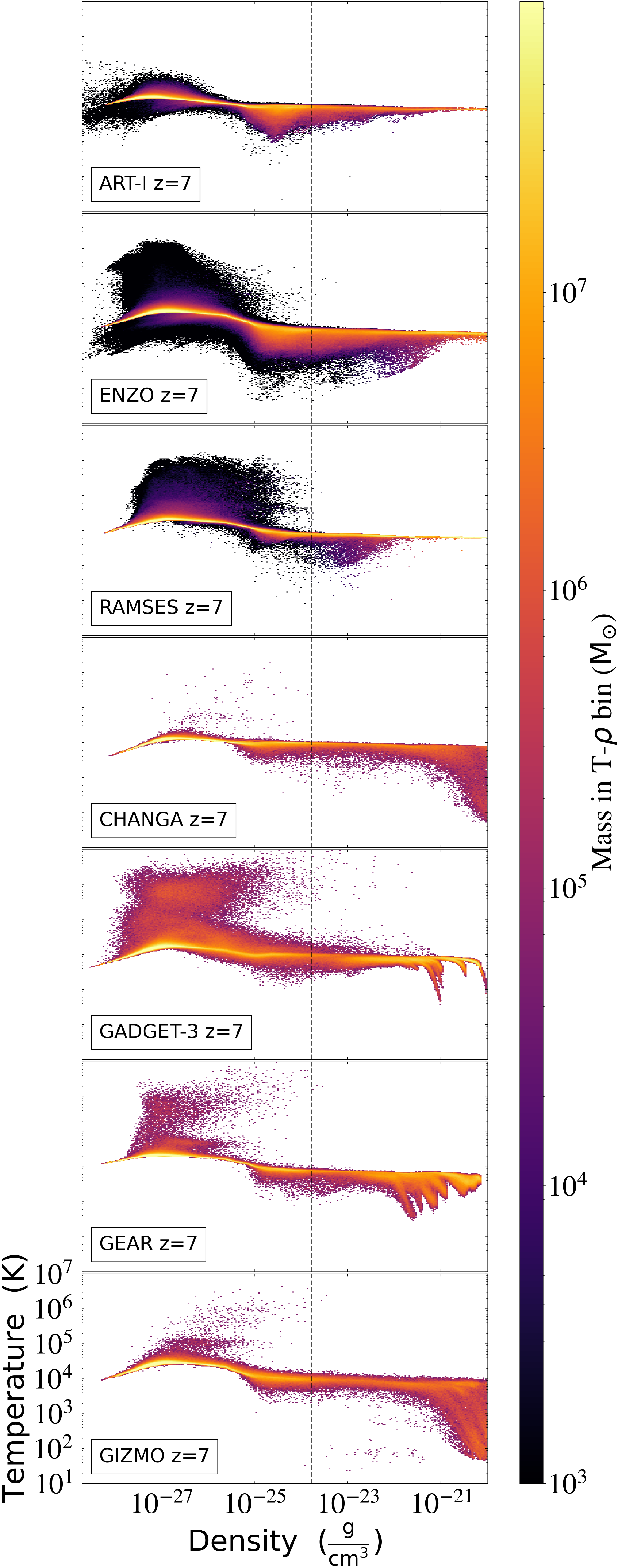}
        \caption{The $z=7$ composite of 2-dimensional PDF of density and temperature for the gas within 100 kpc from the center of the main galactic system in the {\tt Cal-2} runs (cooling and heating test).  
        The 100 kpc-radius sphere encloses the main galaxy, the CGM, and the nearby IGM.  
        Colors represent the total gas mass in each 2-dimensional bin. 
        A black dashed vertical line is placed at the value of the star formation density threshold (Section \ref{common-phy}) to be later adopted in the final simulations in Section \ref{sec:results}. 
        See Section \ref{sec:cal_2} for more information on {\tt Cal-2} and this figure.
        }
        \label{Fig1}
\end{figure}


\subsection{Calibration Step Two \,({\tt Cal-2}):  Cooling and Heating of Gas By Common Physics Package}\label{sec:cal_2}

The second calibration step ({\tt Cal-2}) is designed to check if the common physics package (i.e., cooling, heating, UVB) by {\sc Grackle-v3.1.1} is properly interfaced in all the codes for cosmological runs.
Here, each run is performed with {\sc Grackle-v3.1.1} but {\it without} any subgrid models such as a pressure floor, star formation, or feedback. 
This approach allows us to check the agreement on the gas distribution in the density-temperature plane (expected when the radiative gas physics is treated via the common package  {\sc Grackle-v3.1.1}), and if all codes use the same initial metallicity.

\subsubsection{Findings From {\tt Cal-2}}\label{sec:Cal2_comp}

{\tt Cal-2} has turned out to be a critical calibration step during which the participant code groups found and fixed problems in their {\sc Grackle-v3.1.1} interface.\footnote{During our comparison study using an earlier version of {\sc Grackle}, we found that a small correction on the cooling and heating rates was needed in the {\sc Grackle}/{\sc Cloudy} tables, to ensure correct gas evolution at high redshift.  This issue has been addressed in the {\sc Grackle-v3.1.1} release.}
Note that an earlier version of {\sc Grackle} was implemented and tested for an isolated galaxy disk simulation for all codes (see Section 3.1 of Paper II), but not for a fully cosmological zoom-in run with an expanding simulation volume.   

The gas mass distribution from {\tt Cal-2} in the  density-temperature plane, is shown in Figure~\ref{Fig1} at $z=7$.  
Since the virial radius of the target progenitor at $z=7$ is $\sim7.5$~kpc (see Table~\ref{tab:Generalparams}), and 
we include all the gas inside a sphere of 100 kpc centered on the main halo, the plot includes not just the galactic gas, but most of the IGM inside the Lagrangian zoom-in region. 
Above  $\sim 10^4\,\rm{K}$, the gas cools extremely efficiently owing to both hydrogen and helium recombination. 
Below $\sim 10^4\,\rm{K}$, however, the cooling of the low metallicity primordial gas (see Section~\ref{sec:ic}) is very weak due to the absence of efficient cooling channels other than primordial molecules. 
On the other hand, the low-density gas is strongly heated by the UV background up to $\sim 10^4\,\rm{K}$, while at higher density above the UV self-shielding limit, it is heated by adiabatic compression. 
The combination of these effects leads to the bulk of the gas being found in a well-defined plateau at $\sim 10^4\,\rm{K}$, extending up to high densities ($10^{-20}\,\rm{g\,cm^{-3}}$).

Despite a general good agreement in reproducing this plateau, discrepancies between the participant codes have been noted. 
They reside primarily in the low-density, high-temperature gas in Figure~\ref{Fig1}.
First, it is worth noting that mesh-based codes sample the low-density gas with a large number of bins with {\it small} mass per bin (blue bins) --- which is hard to reproduce by particle-based codes with a (roughly) constant particle mass. 
In Figure~\ref{Fig1}, this leads to a large blue area at density $10^{-27} - 10^{-24}\,\rm{g\, cm^{-3}}$, above and below the $10^4\,\rm{K}$ plateau.  
This area is absent in the particle-based simulations.
Second, a discrepancy exists in the prediction of the rarefied and shocked gas surrounding the halo and filaments.
While particle-based codes predict the presence of the virialized hot gas at $10^{5-6}\,\rm{K}$ (low-density, high-temperature gas around [$\sim10^{-27}\,\rm{g\,cm^{-3}}$, $\sim10^{5-6}\,\rm{K}$] in Figure~\ref{Fig1}, or a similar gas structure in Figure \ref{Fig3} or \ref{Fig5.0}), it is almost absent in the mesh-based codes.
We have carefully studied the behavior of this warm-hot gas, and found that the hot gas is outflowing, while the warm gas is inflowing, confirming that the warm gas surrounding the main galactic system contains shock-heated gas. 
While at this stage of our analysis, the exact origin of the temperature discrepancy between the codes remains unclear, we hypothesize that they result from the different hydrodynamic schemes adopted (differences in the hot virialized gas have already been mentioned in {\tt Cal-1}), and in particular how the schemes treat shocks in strongly cooling gas phases. 

\begin{figure}
        \centering
        \vspace{-3mm}
        \hspace*{-6mm}
        \includegraphics[scale=0.63]{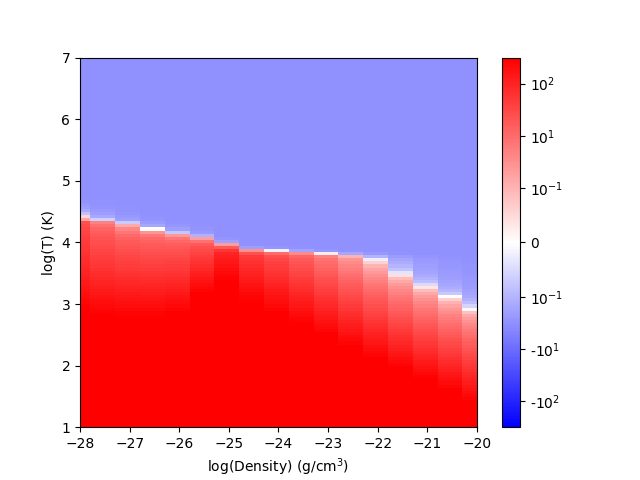}
        \caption{The density and temperature plane colored by the ratio of (heating rate $-$ cooling rate)$/$(cooling rate) in each bin, obtained from the {\sc Cloudy} table at $z\sim7$  in {\sc Grackle-v3.1.1}.}
        \label{Fig2}
        \vspace{0mm}
\end{figure}

Finally, it is worth noting that although those discrepancies may look important, they typically disappear as soon as the stellar feedback is activated (Section \ref{sec:cal_4}). 
Since they have little impact on the star formation in our final {\tt CosmoRun} simulations (Section \ref{sec:results}), we have chosen to defer the detailed discussion to a future paper.   
The extensive studies on the differences in numerical approaches, and how they manifest themselves in the discrepancies in the warm gas surrounding the main galactic system will be in a forthcoming paper by the Collaboration ({\it AGORA} Collaboration et al., in prep.).

\subsubsection{Comments On The Cooling ``Tails'' At High Density In {\tt Cal-2}}\label{sec:CoolTail}

In Figure \ref{Fig1}, a repeating pattern of cooling ``tails'' appears at high density ($\gtrsim 10^{-22}\, {\rm g\,cm}^{-3}$), especially in the particle-based codes {\sc Gadget-3} and {\sc Gear} --- although we have confirmed that these features also exist in {\sc Changa}, {\sc Gizmo}, and the mesh-based codes (e.g., {\sc Ramses}) at other epochs.  
After carefully checking the physics in each of the participant codes, we have found that such features are caused by the cooling and heating tables in our common physics package {\sc Grackle-v3.1.1}. 
To illustrate our finding, in Figure~\ref{Fig2} we show the tabulated rates of primordial cooling and heating at $z\sim7$ from our adopted {\sc Cloudy} table (see Section \ref{common-phy}). 
Here, it is easy to notice how the pre-computed table is binned in density and temperature.
Readers may notice a larger bin size in the density axis, and that the discrete jumps at high density ($\gtrsim 10^{-22}\, {\rm g\,cm}^{-3}$) in the cooling and heating rates exactly coincide with the cooling ``tails'' in Figure~\ref{Fig1}. 
We therefore conclude that the observed cooling ``tails'' originate from the density binning in the pre-computed {\sc Cloudy} table; and, the differences among the participating codes are due to variations in how exactly each code's cooling and heating solver interfaces with {\sc Grackle-v3.1.1}, and its interpolation scheme.

While  the cooling ``tails'' are an interesting observation, we note that these artificial features have little impact on the final cosmological runs presented in Section \ref{sec:results}, because they occur at densities much higher than the star formation threshold, $n_{\rm H, \,thres} = 1\, {\rm cm}^{-3}$, where, in addition, the pressure is dominated by the artificial pressure floor (Section \ref{common-phy}).
The features start to disappear once the dense gas is consumed by stars at later times  (Section~\ref{sec:cal_3}), and will completely vanish as soon as stellar feedback and the pressure floor are activated (Section~\ref{sec:cal_4}).\footnote{Although this feature does not affect the final simulations presented herein, we caution the {\sc Grackle} users when they use the default {\sc Cloudy} tables provided with the package. 
A new table with smaller density bins and/or a careful interpolation scheme would be needed, if interested in studying the very dense gas when no star formation is present.}


\subsection{Calibration Step Three \,({\tt Cal-3}):  Common Star Formation Physics}\label{sec:cal_3}

The third calibration step ({\tt Cal-3}) is designed to detect and study the impact of any discrepancies in the implementation of the common star formation prescription (see Section~\ref{code-phy}). 
Each simulation has been carried out with {\sc Grackle-v3.1.1}, common star formation and pressure floor prescriptions, but without any stellar feedback.
The main objective of {\tt Cal-3} is to ensure that our final cosmological simulation entries in Section~\ref{sec:results} is not dominated by variations (or errors) in how the common star formation physics is implemented in each code. 
At the end of {\tt Cal-3}, each code group confirms that the feedback-free simulations converge within 0.5 dex in stellar masses at $z=7$, and in stellar mass growth history down to that point. 

\begin{figure}
        \centering
        \vspace{2mm}
        \includegraphics[scale=0.25]{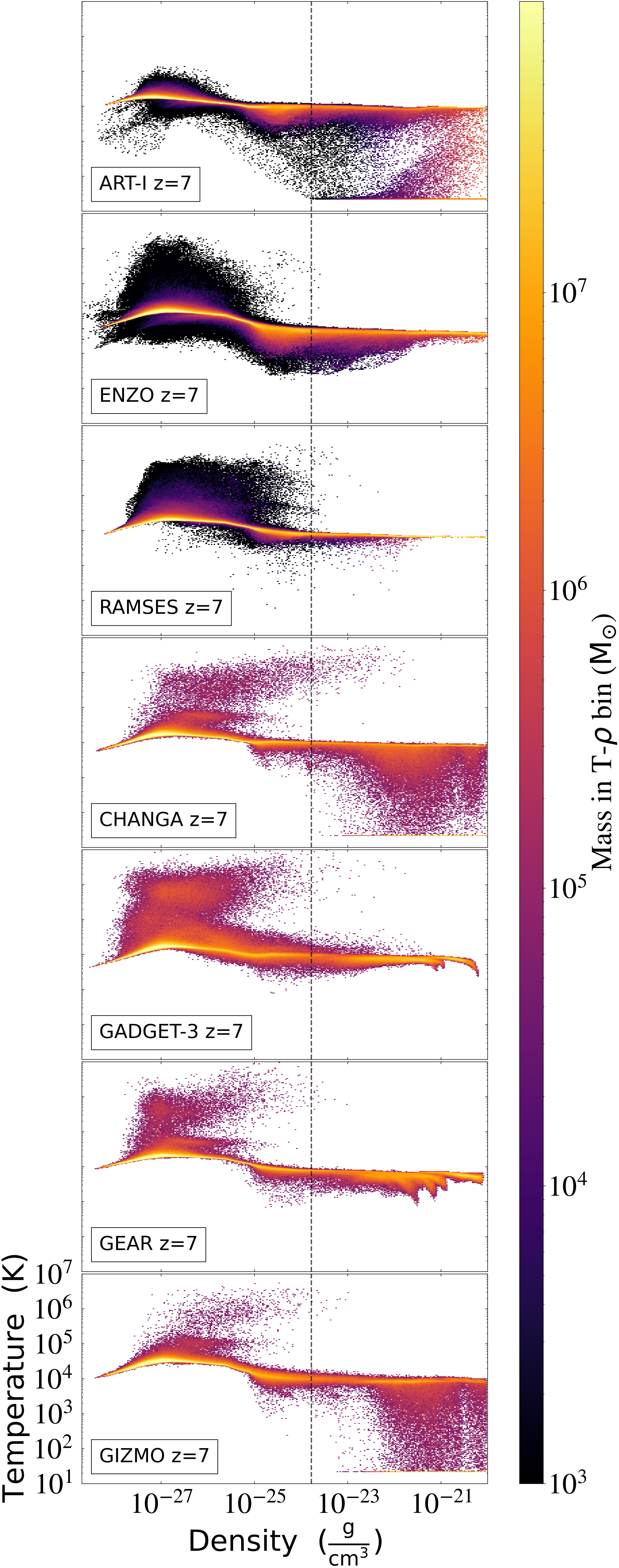}
        \caption{The $z=7$ composite of 2-dimensional PDF of density and temperature for the gas within 100 kpc from the center of the main galactic system in the {\tt Cal-3} runs (star formation test).  
        The 100 kpc-radius sphere encloses the main galaxy, the CGM, and the nearby IGM.  
        Colors represent the total gas mass in each 2-dimensional bin.  
        A black dashed vertical line marks the density threshold for star formation.
        See Section \ref{sec:cal_3} for more information on {\tt Cal-3} and this figure.}
        \label{Fig3}
\end{figure}

\begin{figure}
        \centering
        \vspace{1mm}
        \includegraphics[scale=0.4]{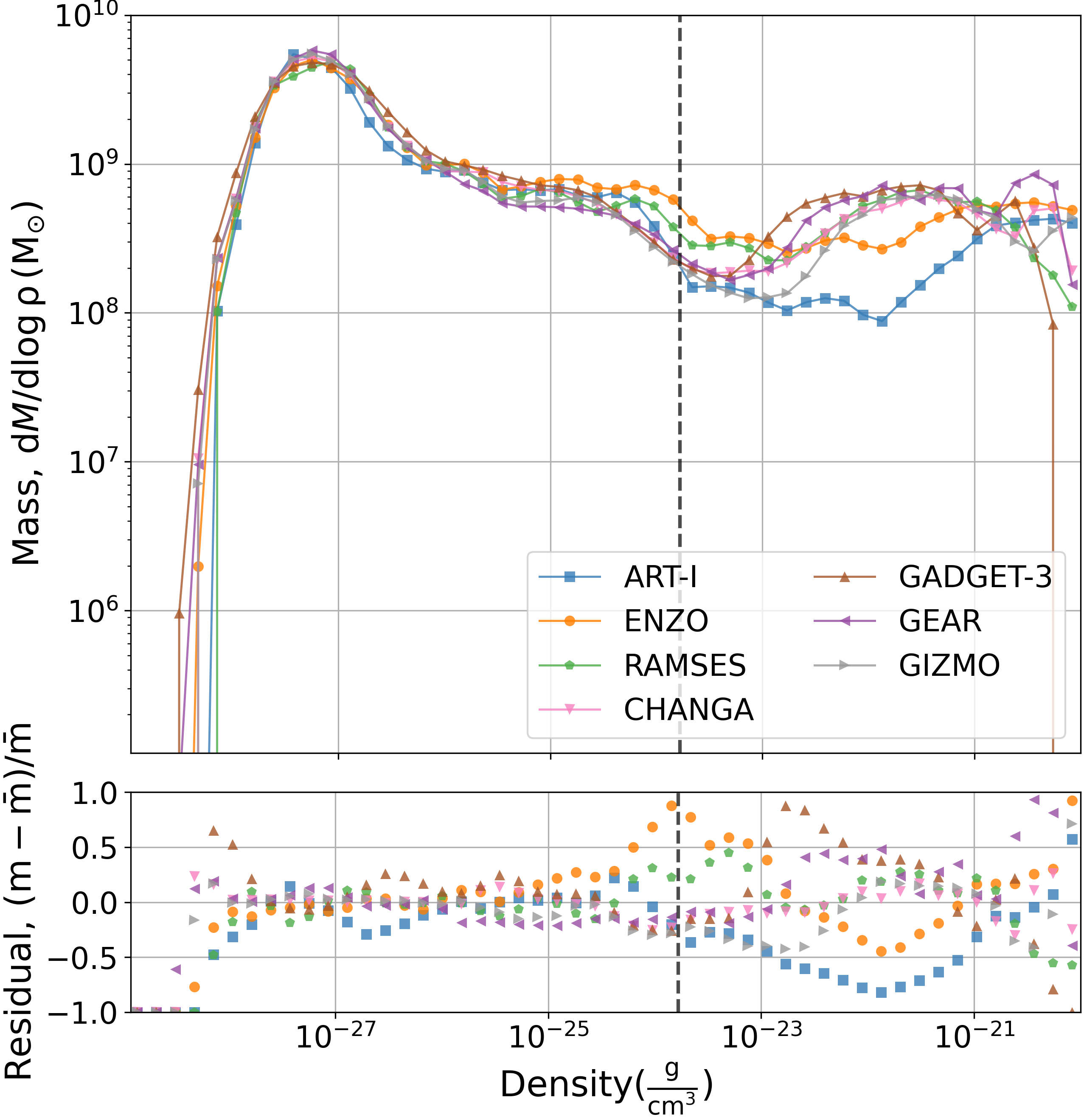}
        \caption{Distribution of gas mass as a function of gas density at $z=7$ for all the gas inside the target progenitor's mean $R_{200}$ ($\sim7.5$~kpc at $z=7$) in {\tt Cal-3}. 
        The vertical dashed line denotes the star formation threshold, $n_{\rm H, \,thres} = 1\, {\rm cm}^{-3}$.  
        Shown in the bottom panel is the fractional deviation from the mean of these profiles.
        See Section \ref{sec:cal_3} for more information on {\tt Cal-3} and this figure.
        }
        \label{Fig3.1}
\end{figure}

\begin{figure}
        \centering
        \vspace{1mm}
        \includegraphics[scale=0.4]{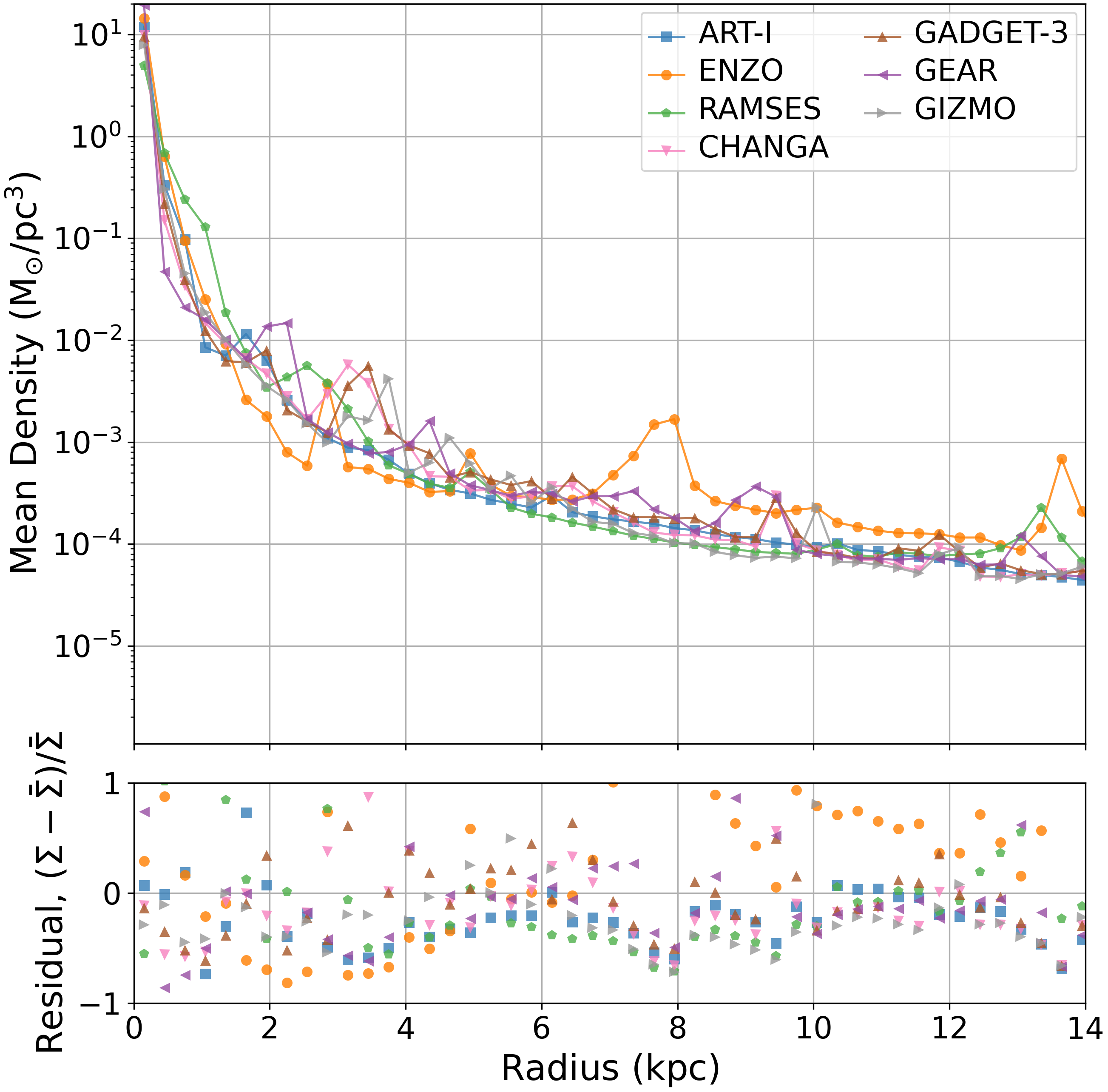}
        \caption{Spherically-averaged gas density profiles as functions of distance from the galactic center at $z=7$ for the {\tt Cal-3} runs. 
        Shown in the bottom panel is the fractional deviation from the mean of these profiles.
        See Section \ref{sec:cal_3} for more information about how the center of the system is selected, the {\tt Cal-3} runs, and this figure.  
        }
        \label{Fig3.2}
\end{figure}

\begin{figure}
        \centering
        \vspace{-3mm} 
        \includegraphics[scale=0.49]{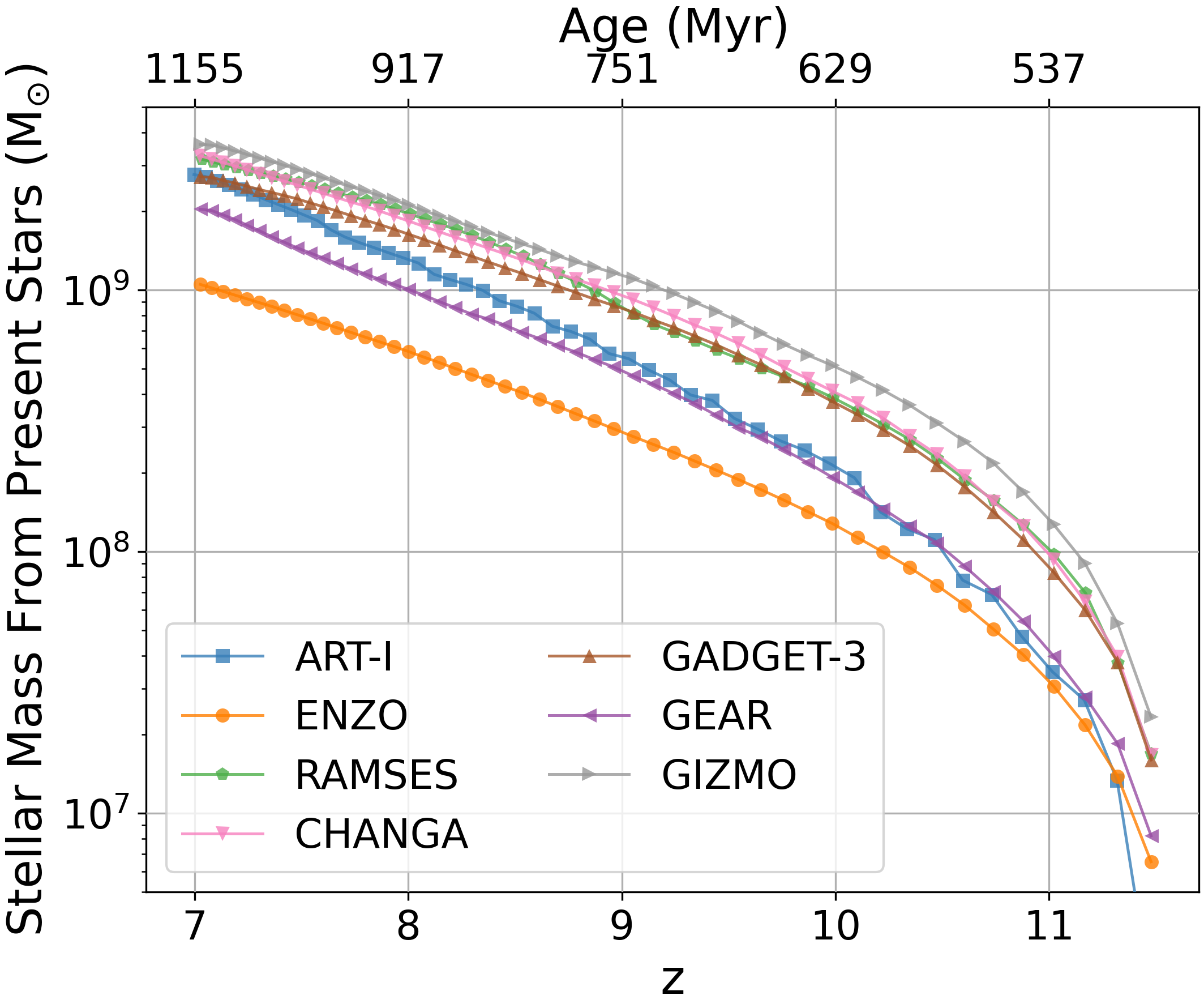}        
        \caption{Stellar mass growth histories for the {\tt Cal-3} runs in a 100 kpc sphere centered at the target progenitor. 
        The curve is computed using the ages or creation times recorded in star particles at $z=7$.}
        \label{Fig4}
        \vspace{1mm}
\end{figure}

\begin{figure*}
        \centering
        \vspace{0mm}
	\includegraphics[scale=0.6]{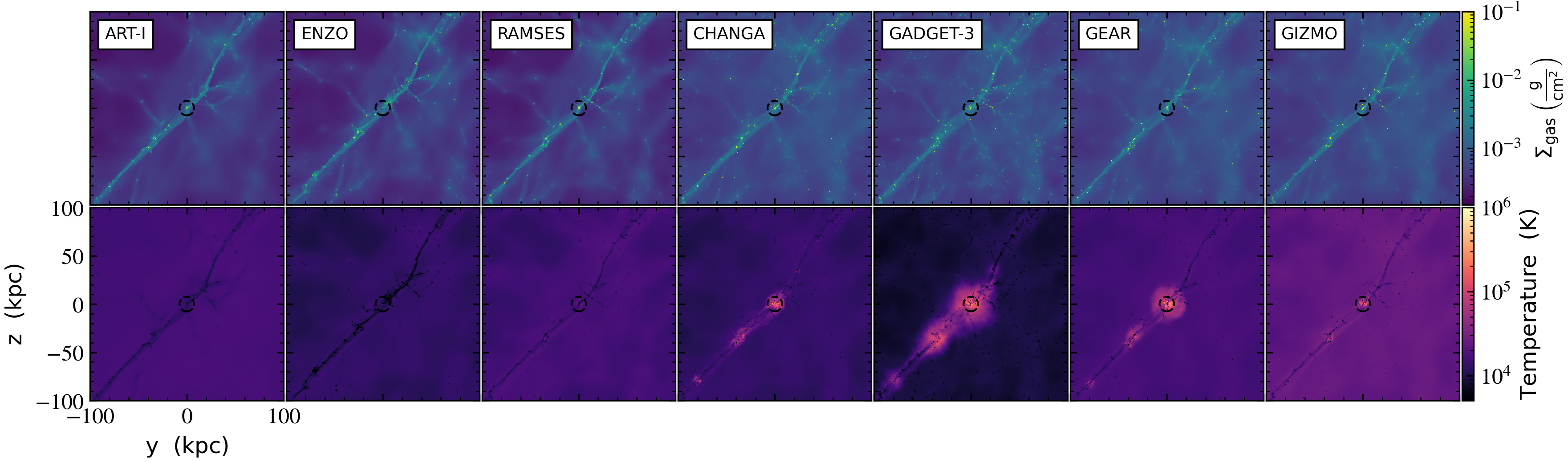}
        \caption{Gas density projection ({\it top}) and density-weighted temperature projection ({\it bottom}) at $z=7$ from the third calibration step, {\tt Cal-3} (star formation test). 
        We indicate the mean $R_{200}$ among the codes ($\sim 7.5$ kpc) with a black dashed circle. 
        Units are proper kpc.  
	The projections along the other axes are available as digital supplements to this article.  
	See Section \ref{sec:cal_3} for more information on {\tt Cal-3} and this figure. 
        }
        \label{Fig5.0}
        \vspace{1mm}
\end{figure*}

\begin{figure*}
        \centering
        \vspace{0mm}
        \includegraphics[scale=0.6]{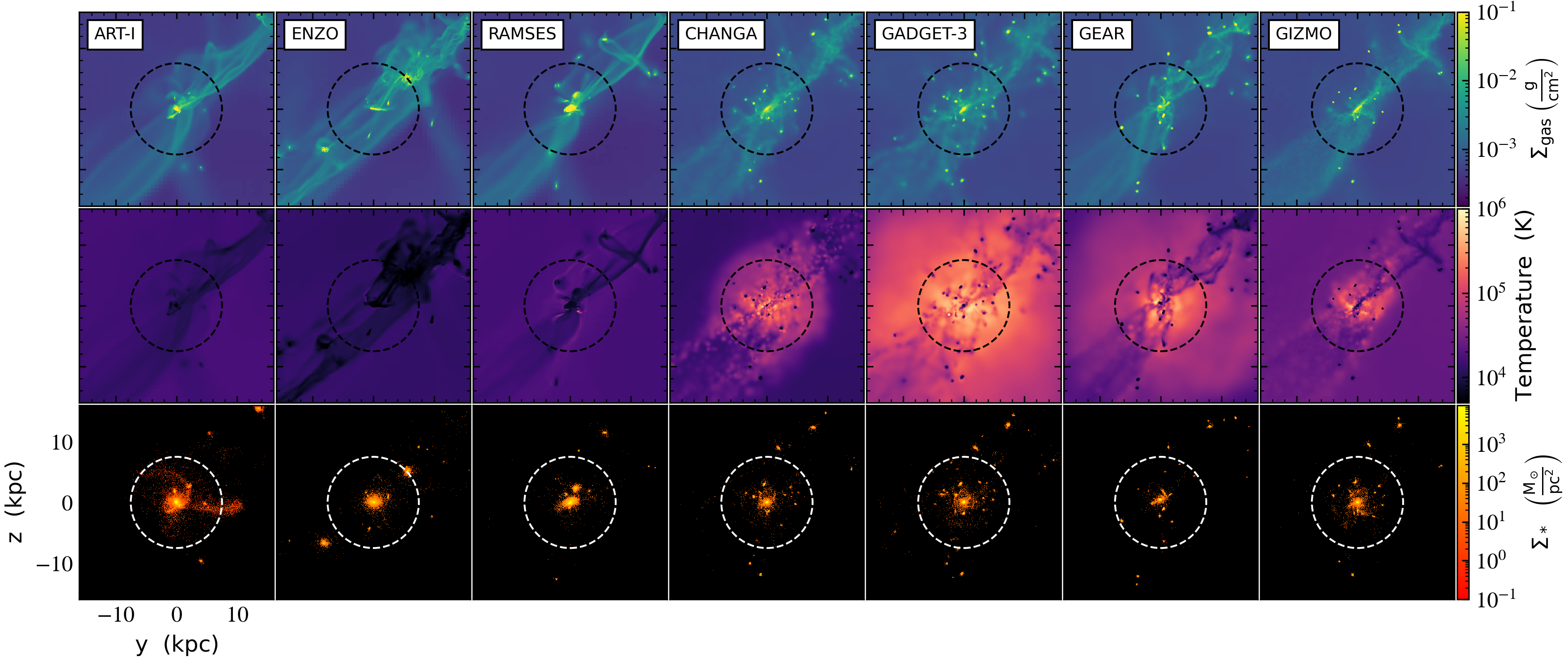}
        \caption{Similar to Figure \ref{Fig5.0}, but now in zoomed-in regions.  Gas density projection ({\it top}), density-weighted temperature projection ({\it middle}), and stellar surface density ({\it bottom}) at $z=7$ from the third calibration step, {\tt Cal-3}. 
        The width of each panel is $4R_{200} = 30 \,{\rm kpc}$. 
        The mean $R_{200}$ among the codes ($\sim 7.5$ kpc) is indicated with a black/white dashed circle.         
        See Section \ref{sec:cal_3} for more information on {\tt Cal-3} and this figure.
	}
        \label{Fig5}
        \vspace{3mm}
\end{figure*}

\subsubsection{Findings From {\tt Cal-3}}\label{sec:results_cal3}

In Figure~\ref{Fig3}, we plot  the 2-dimensional density-temperature PDF at $z=7$. 
It displays a good agreement on the general features in the density-temperature plane, such as the shape of the $\sim 10^4$ K cooling plateau where most of the gas mass resides. 
Nevertheless, there are differences, some of which were discussed in previous sections --- e.g., a large number of bins with small mass  in the low-density, high-temperature region (blue bins; Section~\ref{sec:Cal2_comp}), and the cooling ``tails'' at high density (Section~\ref{sec:CoolTail}).
An interesting new discrepancy in Figure~\ref{Fig3} is the presence of high-density, low-temperature gas found in {\sc Art-I}, {\sc Changa} and {\sc Gizmo}, with its density near the star formation threshold and its temperature near the CMB floor. 
This artificial feature results from using a stochastic star formation recipe and a particular pressure floor implementation (see Sections~\ref{sec:code-art} and \ref{sec:code-changa});\footnote{The {\sc Art-I} code, for example,  uses stochastic star formation along with a treatment to avoid complete gas depletion in a star-forming gas cell (see Section~\ref{sec:code-art}). 
Hence, after a cell spawns a star particle, a fraction of gas is still left in the cell with the same temperature as before but with a significantly lowered density. 
Due to the imposed pressure floor, the equilibrium with the surrounding cells can only be achieved through rapid cooling, and a slightly increase on the density. This process results in a build-up of the observed cold gas near the CMB floor. 
Similar features have been reproduced  in other codes (e.g., {\sc Ramses}) when stochastic star formation is employed. \label{footnote:art-i-stochastic}}
however, the discrepancy becomes largely marginal once stellar feedback is turned on as we will discuss it in  Section \ref{sec:cal_4} and Figure \ref{Fig13}.

Nevertheless, on the whole, the {\tt Cal-3} entries from the participating code groups exhibits robust overall convergence in the gas distribution around the target progenitor galaxy, as illustrated in Figures \ref{Fig3.1} to \ref{Fig4}. 
In Figure~\ref{Fig3.1}, we display the gas mass distribution as a function of its density, including all the gas inside the virial radius $R_{200}$ ($\sim 7.5$ kpc at $z=7$). 
We find that all participant codes produce a very similar gas density probability distribution inside $R_{200}$. 
Note that the convergence is better than in our disk comparison (Figure 18 of Paper II) in which, by design, gaseous halos --- low-density tails towards the left side of this plot --- existed only in mesh-based codes, but not in particle-based codes. 
In Figure~\ref{Fig3.2}, we show the spherically-averaged gas density as a function of radius, again demonstrating solid convergence aside from small variations due to the halo substructures and clumps.\footnote{The profile center  is set to be the location of maximum stellar density within a successively shrinking distance from the dark matter center of mass.}  
In both Figures 7 and 8 we include the fractional deviation from the mean of these profiles to better illustrate the convergence among the codes.

The most relevant result from {\tt Cal-3} is, however, the convergence in the stellar mass $M_\star$ evolution (in a 100\,kpc sphere centered at the target progenitor) in Figure~\ref{Fig4}. 
Though small variations exist, all codes follow similar stellar mass growth histories, within half a dex from one another at all times. 
Differences among codes are due to variations in how the common star formation prescription is implemented (e.g., stochastic in {\sc Art-I}, {\sc Changa}, {\sc Gadget-3}, {\sc Gear} vs. deterministic in {\sc Enzo} and {\sc Ramses}; see Section~\ref{common-phy}), refinement strategy (Section~\ref{sec:runtime}), and/or numerical accuracies of hydrodynamics solvers (Section 5 of Paper II).\footnote{Note that {\sc Enzo} produces 2-3 times fewer stars than the other codes. 
Unlike the other codes, the only tunable parameters in {\sc Enzo}'s star formation module is the star formation efficiency and the density threshold, both being fixed in this work (see Section~\ref{common-phy}).  
Thus, it has been difficult to further adjust {\sc Enzo}'s star formation to acquire better convergence.}

\subsubsection{Comments On The Differences In Galactic Morphology In {\tt Cal-3}}

Finally, a detailed comparison of the gas and stellar distribution in real space is shown in Figure~\ref{Fig5.0} and \ref{Fig5}. 
In Figure~\ref{Fig5.0} we show the projected gas density (top row) and temperature (bottom row) of all the gas inside the $(200 \,\, {\rm kpc})^3$ volume (compare with Figure \ref{Fig00} in {\tt Cal-1}). 
The mean virial radius $R_{200}$ among the codes is shown as a black dashed circle. 
In the gas density map, the  large-scale structures are nearly identical across all participant codes, although the aforementioned differences in the low-density region (discussed in Section \ref{sec:Cal1_comp} and \ref{sec:Cal2_comp} with {\tt Cal-1} and {\tt Cal-2}, respectively) still exist between the mesh-based and particle-based approaches.  
Figure~\ref{Fig5} demonstrates this more dramatically, in which we show the projected gas density (top row), temperature (middle row), and stellar surface density (bottom row) at $z=7$ inside a $(4R_{200})^3$ volume.
Notable is that,  in the stellar surface density map, the particle-based codes harbor more satellites (clumps of star particles) than the mesh-based codes. 
This discrepancy is caused by the same effect that leads particle-based codes to preserve more substructures in the low-density region.  
It has been well documented that due to the lack of force resolution at high $z$, mesh-based codes tend to suppress the low-mass end of the halo mass function (see Section \ref{sec:Cal1_comp} of this article, or Section 5.3.2 in Paper II).  

Also in Figure~\ref{Fig5}, differences exist in the temperature map between the mesh-based and particle-based codes, particularly in the regions next to the galaxies and filaments. 
This difference manifests itself as a diverging distribution in the density-temperature PDF near $\sim 10^{-27}\,{\rm g\,cm^{-3}}$, $\sim 10^{5-6}$ K in Figure \ref{Fig3}. 
We recall, however, from Section \ref{sec:Cal2_comp} that the observed temperature differences become irrelevant as soon as the stellar feedback is activated, and thus have little impact on the results of the final zoom-in cosmological runs  ({\tt CosmoRun}) in Section \ref{sec:results}.


\begin{figure}
        \centering
        \vspace{1mm}
        \includegraphics[scale=0.49]{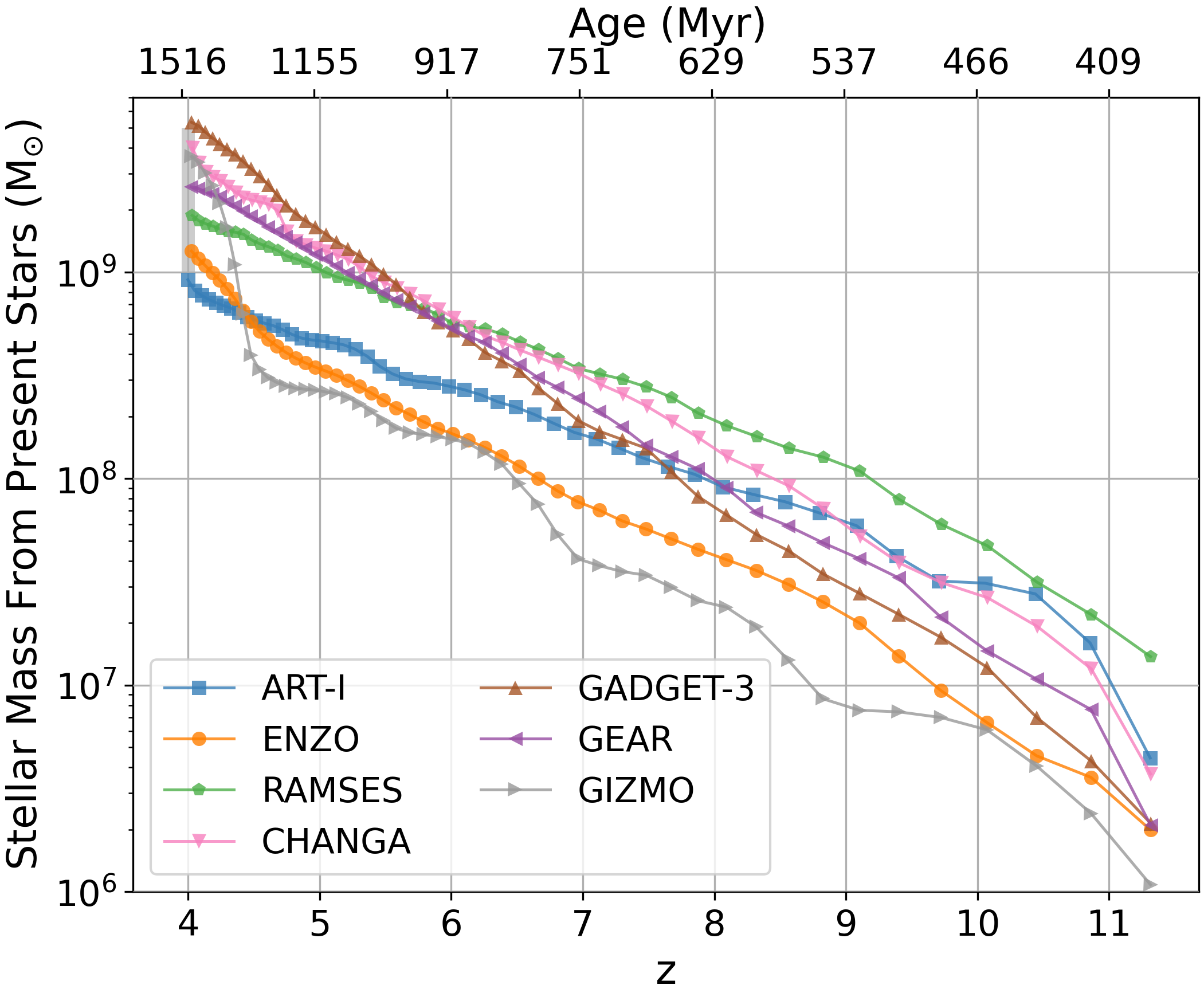}
        \caption{Stellar mass growth histories for the {\tt Cal-4} runs inside a $R_{200}$ sphere centered at the target progenitor. 
        The curve is computed using the ages or creation times recorded in star particles at $z=4$.
        The stellar mass range at $z=4$ targeted in our calibration  is $M_\star \sim 1-5\times 10^9\, {\rm M}_{\odot}$, as motivated by semi-empirical models. 
        What we show here is an upper limit for the total $M_\star$ formed inside $R_{200}$. 
        It is in Figure~\ref{Fig10} where we can make a fair comparison of $M_\star$ formed inside the galaxy with predictions from semi-empirical models.
        See Section \ref{sec:cal_4} for more information on {\tt Cal-4} and this figure.
        }
        \label{Fig6}
        \vspace{0mm}        
\end{figure}

\subsection{Calibration Step Four \,({\tt Cal-4}):  ``Favorite'' Stellar Feedback Prescription By Each Code}\label{sec:cal_4}

The objective of this last calibration step  ({\tt Cal-4}) is to get convergence on the stellar mass of the main progenitor at $z=4$ within 0.5 dex, to the value predicted by the semi-empirical models based on the abundance matching techniques \citep[e.g.,][]{Rodriguez_puebla2017}. 
The main motivation for {\tt Cal-4} is to come up with a {\it realistic} simulation resembling observed galaxies, by adopting each code group's ``favorite'' feedback --- as close to the most widely-used one for research in each code community.  
Each code group's cosmological simulation has been carried out with {\sc Grackle-v3.1.1}, a common star formation prescription, and its own choice of stellar feedback and metal production (see  Table \ref{tab:feedback} and Section \ref{code-phy}).
Each group has been asked to provide a reference with detailed information on their ``favorite'' feedback prescription (as in Section \ref{code-phy}).
Although time-consuming, at the end of {\tt Cal-4} we establish a common ground based on which we can compare  the effects of each group's ``favorite'' feedback on the evolution of galaxies and CGM.

\subsubsection{Calibration Target In {\tt Cal-4}}\label{sec:target_cal4}

According to the predictions by the aforementioned semi-empirical models, the expected stellar mass inside the main galactic system of a $M_{200}=2\times10^{11}\, {\rm M}_{\odot}$ halo at $z=4$ is $\sim 1-1.5\times 10^9\, {\rm M}_{\odot}$. 
Since our selected halo (see Section \ref{sec:ic}) experiences a relatively violent assembly history by $z=4$, we have extended the target range of the stellar mass $M_\star$ to $\sim 1-5\times 10^9\, {\rm M}_{\odot}$ at $z=4$. 
The width of the target mass range is  to allow  flexibility when each code group selects its stellar feedback scheme.  
{\tt Cal-4} has required the most amount of time among all calibration steps.   
Typically, the process was not over with a single simulation, but required several iterations carried out by each participating code group.  
The simulations they acquire after these iterations become the final entries in Section~\ref{sec:results} (dubbed {\tt CosmoRun}).
In this subsection we briefly discuss only the calibration process in {\tt Cal-4}, not the detailed analysis of each code group's final simulation entry --- the latter will be discussed in full detail in  Section~\ref{sec:results}.

\subsubsection{Findings From {\tt Cal-4}}\label{sec:results_cal4}

At the end of {\tt Cal-4}, the participating code groups have found a need to use stronger stellar feedback than they commonly used in their communities in order to achieve the target stellar mass at $z=4$. 
However, none of them used unrealistic feedback parameters.  
In Figure~\ref{Fig6} we show the stellar mass growth histories of final simulation entries. 
Each curve has been obtained using the star particles residing inside a $R_{200}$ sphere centered on the target progenitor galaxy at $z=4$.
Therefore, Figure~\ref{Fig6} is the stellar mass assembly history (SMAH) inside $R_{200}$, not the star formation history (SFH) of the main galactic system, thus it is only an upper limit for the generated stellar mass.\footnote{Unlike the SFH, the SMAH includes not only the stars formed inside the target progenitor (in-situ), but also the stars formed outside and brought in by e.g., merging satellites (ex-situ).  In the SMAH, the stellar mass may decrease due to the mass loss when the galaxy interacts with its neighbors.
In future studies, we plan to compare the actual SFH (rather than SMAH). \label{footnote:mah}} 
The plot demonstrates how all codes successfully converge to the agreed $M_\star$ range, although the SPH codes tend to have higher $M_\star$ at $z=4$. 
Comparing Figure~\ref{Fig6} with Figure~\ref{Fig4}, in each code we observe the expected decrease of the stellar mass growth due to the stellar feedback (notice the change in the $y$-axis). 
The shape of the SMAH differs from one code to another because of the different stellar feedback prescriptions implemented in the codes, that can affect star formation differently at a given epoch.
The ``timing discrepancies'' among the codes in the halo assembly history could also cause differences in the SMAHs.
Indeed, the exact timing of a major merger occurring at $z\sim4$ could precipitate sizable variations in the SMAH, and the gas and stellar properties discussed in Section \ref{sec:results} (see Sections \ref{sec:CosmoRunGas} and \ref{sec:CosmoRunStars} for more discussion).\footnote{The discrepancies in the exact timings of mergers and star formation events, could affect the discussion of various galactic properties in Section \ref{sec:results}.  
In particular, at high $z$, major mergers are common and can violently disturb the gas inside the galaxy and in its CGM by generating shocks and changing the gas distribution in the density-temperature plane. 
These perturbative events do not occur at the exact same redshift in all codes (see Section 5.3.2 of Paper I), complicating the inter-code comparison. 
In the future papers, we will extensively study variations in the participating codes' merger trees.\label{footnote:timing}}
Lastly, readers may notice that the inter-code differences are larger at early times (e.g., the variation is $\sim1.5$ dex at $z=10$ but $\sim0.5$ dex at $z=4$). 
Indeed, previous research have found that different stellar feedback implementations can exacerbate the discrepancy at high redshift \citep[e.g.,][]{2017MNRAS.465.1682H}. 

\begin{figure}
        \centering
        \vspace{3mm}
        \includegraphics[scale=0.60]{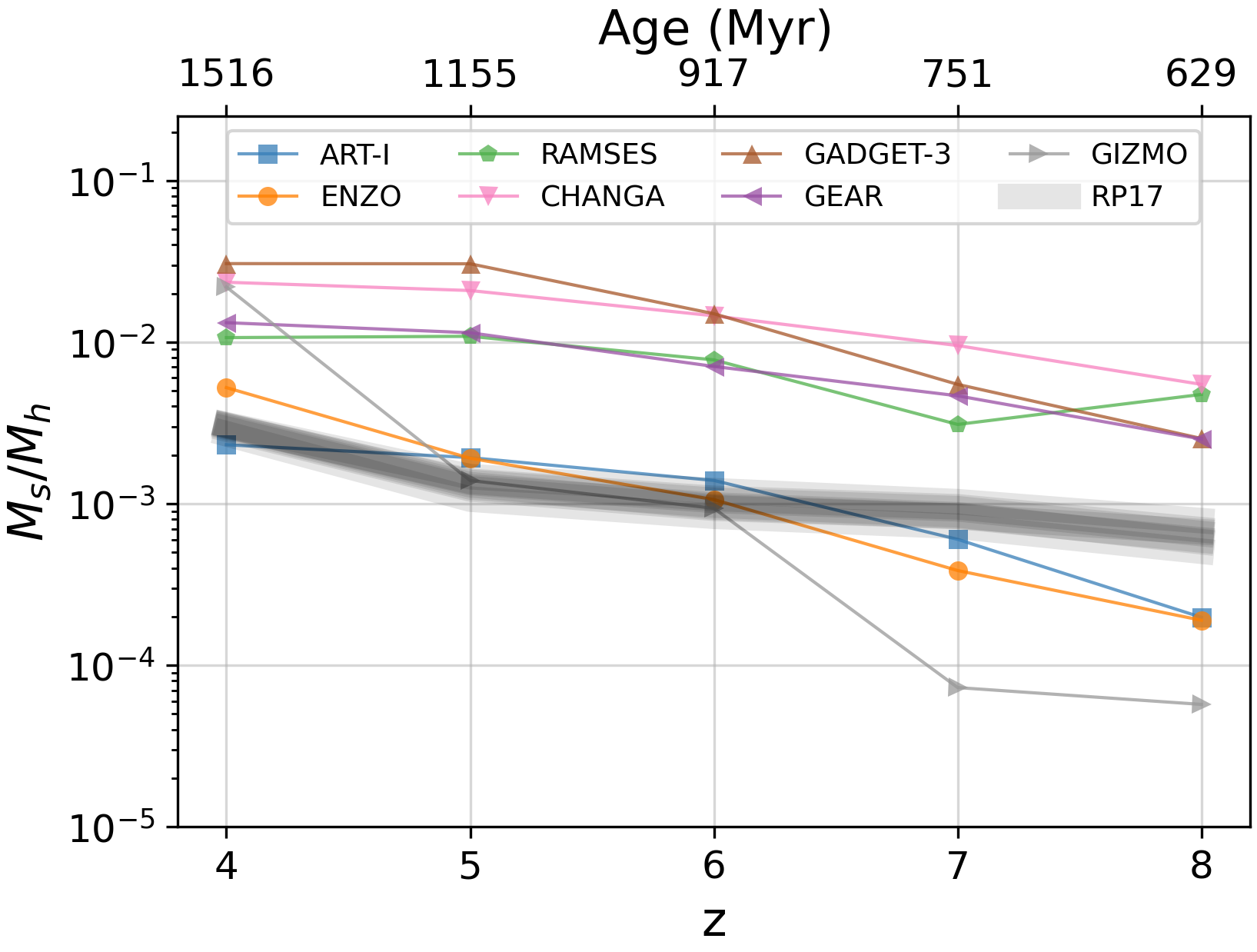}
        \caption{Evolution of the stellar-to-halo mass ratio, $M_{\rm \star, \, gal} / M_{200}$, from $z=8$ to $z=4$ in the {\tt CosmoRun} simulations (rightmost column in Table \ref{tab:Generalparams}). 
        Gray shadowed regions indicate the predicted ranges of the ratio by the semi-empirical model of \citet{Rodriguez_puebla2017}, obtained using the halo mass at each redshift, in each simulation.
        See Section \ref{sec:results} for more information on {\tt CosmoRun}, and Section \ref{sec:CosmoRunGlobal} in particular on this figure.
        }
        \label{Fig10}
        \vspace{0mm}
\end{figure}

\vspace{3mm}

With this final result, we conclude the entire calibration procedure. 
The code groups that completed the four calibration steps, {\tt Cal-1} to {\tt Cal-4}, have obtained the final {\tt CosmoRun} simulations. 
In the next Section, we present and analyze the properties of these final simulation entries  from the codes groups down to $z=4$. 

\begin{table*}
\footnotesize
\vspace{2mm}
\caption{\footnotesize Global properties of the target galaxy progenitor in the {\it AGORA} {\tt CosmoRun} simulation suite}
\centering
\npdecimalsign{.}
\nprounddigits{2}
\begin{tabular}{c c || cccccc}
\hline\hline
Code & redshift $z$ & $M_{200}^{\,\,\,(a)}\tablenotemark{\textdagger}$ & $M_\star^{\,\,\,(b)}$ & $M_{\rm gas}^{\,\,\,(c)}$ & $M_{\rm gas, \,gal}^{\,\,\,(d)}$ & $M_{\rm gas, \,CGM}^{\,\,\,(e)}$ & log$(M_{\rm \star, \, gal} / M_{200})^{\,(f)}$  \\
& & [10$^{10}\,$M$_{\odot}$] & [10$^8\,$M$_{\odot}$] & [10$^8\,$M$_{\odot}$] & [10$^8\,$M$_{\odot}$] & [10$^8\,$M$_{\odot}$] & \\
\hline
{\sc Art-I} & 8   &  0.92   &  0.48   &  11.36 &  0.19  &  11.18   &  -3.7  \\
&7   &  1.49   &  1.04   &  14.87   &  0.38  &  14.50   &  -3.22    \\
&6   &  1.83   &  1.52   &  17.80  &  0.56   &  17.24   &  -2.86     \\ 
&5   &  2.77   &  1.98   &  28.50   &  1.29   &  27.21   &  -2.71     \\
&4   &  13.23   &  9.22   &  145.41   &  21.68  &  123.72   &  -2.64  \\
\hline
{\sc Enzo} & 8    &  1.16   &  0.23   &  11.03  &  0.17   &  10.86   &  -3.72  \\
&7    &  1.84   &  0.43   &  22.37  &  0.83  &  21.54  & -3.41   \\
&6   &  2.26   &  0.96   &  30.05 &  1.58  &  28.46  &  -2.97   \\
&5    &  3.84   &  2.04   &  51.41   &  3.67  &  47.74   &  -2.72   \\
&4    &  16.04   &  12.72   &  242.62  &  58.39  &  184.23   &  -2.28   \\
\hline
{\sc Ramses} & 8   &  1.37   &  1.21   &  17.73   &  2.97  &  14.75  &  -2.32   \\
&7   &  1.84   &  1.67   &  19.85   &  1.51   &  18.35   & -2.51   \\
&6   &  2.19   &  2.87   &  26.59  &  5.12  &  21.48  &  -2.11   \\
&5   &  3.50   &  5.12   &  36.51  &  10.43   &  26.08    &  -1.96  \\
&4   &  14.79   &  18.98   &  139.47   &  44.32  &  95.15   &  -1.97   \\
 \hline
{\sc Changa} & 8    &  1.43   &  1.17   & 29.03   &  5.94   &  23.37   &  -2.26  \\
&7    &  2.26  &  2.82   &  43.22  &  7.55  & 35.67  &  -2.02  \\
&6   &   2.72  &  5.09   &  58.88 &  17.91  & 40.97   &  -1.84   \\
&5    &  4.15   &  10.89   &  72.74   & 11.76   & 60.98    & -1.68    \\
&4    &  15.81   &  39.94   &  203.04  &  85.70  &  117.34   &  -1.63  \\
 \hline
{\sc Gadget-3} & 8    &  1.32   &  0.48   &  25.16   &  5.62   &  19.54  &  -2.60   \\
&7    &  2.17   &  1.47   &  38.84  &  7.41   &  31.43  &  -2.26  \\
&6    &  2.61   &  4.23   &  49.25   &  18.06 &  31.20   &  -1.82   \\
&5    &  4.05   &  12.75   &  71.65  &  26.46   &  45.20   &  -1.52  \\
&4    &  16.15   &  53.17   &  216.98  &  76.24  &  140.74  &  -1.51  \\ 
\hline
{\sc Gear} & 8   &  1.72   &  0.67   &  39.52  &  8.28  & 31.24  &  -2.60   \\
&7   &  2.52   &  1.55   &  58.84  &  15.51  &  43.33  &  -2.33   \\
&6   &  3.23   &  3.71   &  82.14  &  14.93  &  67.21 &  -2.15    \\
&5   &  4.60   &  7.77   &  111.38  &  40.51  &  70.87  &  -1.94   \\
&4   &  16.34   &  25.92 &  286.33  &  145.52  &  140.81  &  -1.88  \\
\hline
{\sc Gizmo} &8    &  1.12   &  0.14   & 10.96   &  0.0  & 10.96 &  -4.24   \\
&7    &  1.90   &   0.20  &  24.56  &  1.15  & 23.41 &  -4.14     \\
&6    &  2.35   &  0.92   &  33.02  &  0.98  & 32.04 & -3.03       \\
&5    &  3.65   &  1.64   &  41.18  &  1.32  & 39.86 &  -2.86     \\
&4    &  15.39   &  36.23   &  165.59  &  41.21  & 124.38 &  -1.66    \\
\hline
\end{tabular}
\tablenotetext{$\textdagger$}{\scriptsize Each column lists the following quantities at the corresponding redshift: $^{(a)}$total halo mass, $^{(b)}$stellar mass, $^{(c)}$gas mass inside the mean $R_{200}$ among codes, where the  $R_{200}$ values found are 5.8, 7.5, 8.4, 11.4 and 25.4 proper kpc at $z=8, \, 7, \, 6, \, 5$ and 4, respectively, $^{(d)}$gas mass inside the main galaxy or the ISM (which we define as regions with $R < 0.15 \,R_{200}$), $^{(e)}$gas mass in the CGM  (which we define as regions with $0.15 \,R_{200} < R < R_{200}$),  $^{(f)}$the ratio of stellar mass (in the main galaxy) to halo mass.}
\label{tab:Generalparams}
\vspace*{1mm}
\end{table*}

\vspace{5mm}

\section{The {\it AGORA} \texorpdfstring{\tt C\MakeLowercase{osmo}R\MakeLowercase{un}}{CosmoRun} {\rm Simulations}} \label{sec:results}

\vspace{1mm}

In this section, we introduce the {\it AGORA} {\tt CosmoRun} simulations acquired from the rigorous calibration steps in Section \ref{sec:calibration}.
As we present the analysis of their stellar and gas components, we focus on five redshifts, $z=8,\,7,\,6,\,5$ and $4$.\footnote{$1.09$, $1.22$, $1.40$, $1.63$, $1.96\,\,\rm{Gyr}$ in cosmic time, respectively.}
The simulations have been running down to even lower redshift, and the full analysis --- the CGM evolution down to e.g., $z=2$, in particular ---  will be presented in the forthcoming papers from the {\it AGORA} Collaboration.

\vspace{1mm}

\subsection{Global Properties of The Target Galaxy Progenitor}\label{sec:CosmoRunGlobal}

We start by analyzing the global bulk properties of the target galaxy progenitor in {\tt CosmoRun}. 
In Table~\ref{tab:Generalparams} we list the total virial mass, $M_{200}$, and gas and stellar masses enclosed inside a sphere whose radius is  the mean $R_{200}$ among the codes. 
We also include the gas masses inside the main galaxy vs. those in the CGM (i.e., $M_{\rm gas, \,gal}$ for $R < 0.15 \,R_{200}$ vs. $M_{\rm gas, \,CGM}$ for $0.15 \,R_{200} < R < R_{200}$), and the stellar-to-halo mass ratio, $M_{\rm \star, \, gal} / M_{200}$, obtained by using the star particles inside $0.15\, R_{200}$ (rightmost column in Table~\ref{tab:Generalparams}; see also Figure~\ref{Fig10}).
It should be noted that we do not expect to find perfect convergence in all the properties here, but expect substantial dependence on the stellar feedback prescriptions adopted by each code group. 
This dependence will be especially evident in the spatial distribution of gas in and around the target halo, and also in its temperature and metallicity. 

Table~\ref{tab:Generalparams} illustrates that all the participating codes converge on the stellar and total masses within $<$ 0.5 dex from one another. 
This convergence is not surprising as it is a consequence of the calibration strategy used ({\tt Cal-4}; Section~\ref{sec:target_cal4}). 
The small deviations from code to code in the total mass, $M_{200}$, are due to the ``timing discrepancies''  in the halo assembly history (Section \ref{sec:results_cal4} and footnote \ref{footnote:timing}). 
On the other hand, relatively larger deviations in the gas mass inside the virial radius, $M_{\rm gas}$, or the ratio of gas masses in the main galaxy vs. in the CGM (i.e., $M_{\rm gas, \,gal}$ vs. $M_{\rm gas, \,CGM}$), are a direct consequence of the different stellar feedback strategies adopted. 
In fact, the strength of the outflows generated by stellar feedback has a strong impact not only on the amount of gas remaining inside the virial radius, but also on how efficiently the cold inflows replenish the galaxy with fresh gas. 
A detailed analysis of the thermodynamics and kinematics of gas is in Sections~\ref{sec:CosmoRunGas} and \ref{sec:CosmoRunCGM}.

In Figure~\ref{Fig10} we show the stellar-to-halo mass ratios, $M_{\rm \star, \, gal} / M_{200}$, in the {\tt CosmoRun}, computed at $z=8, \,7, \,6, \,5$ and 4 (see also the rightmost column in Table~\ref{tab:Generalparams}),  compared with predictions from semi-empirical models \citep[e.g.,][]{Rodriguez_puebla2017}. 
The gray shadowed regions indicate the stellar-to-halo mass ratio obtained from a semi-empirical model using $ M_{200}$ at each redshift, in each simulation. 
Since in {\tt Cal-4} we calibrated each simulation's stellar feedback so that the stellar mass produced is in the range of $\sim 1-5\times 10^9\, {\rm M}_{\odot}$ at $z=4$ (see Figure \ref{Fig6} and Section \ref{sec:target_cal4}), all seven lines do not deviate more than one dexfrom one to another at $z=4$. 
In addition, the difference between the simulated stellar-to-halo mass ratios and the semi-empirical predictions is less than 1~dex at $z=4$, because it is designed as such in {\tt Cal-4}.
However, the semi-empirical predictions lie below the simulated values in most codes.
The mismatch is because our target halo does not have an assembly history of a prototypical halo of $10^{12} \,{\rm M}_{\odot}$ at $z=0$, but that of a halo which assembled early and had a quiescent period from $z=2$ to 0 (Section \ref{sec:ic}).  
This bias yields a higher-than-expected stellar mass at $z\gtrsim 4$.   
At higher redshift $(z \gtrsim 7)$, the differences among the simulated stellar-to-halo mass ratios, and that between the simulated ratios and the semi-empirical predictions are significantly larger.  
They are due to the variations in the feedback prescriptions, causing changes on the amount of star-forming gas available at each redshift, hence on the star formation history.

\begin{figure}
        \centering
        \vspace{2mm}        
        \includegraphics[scale=0.25]{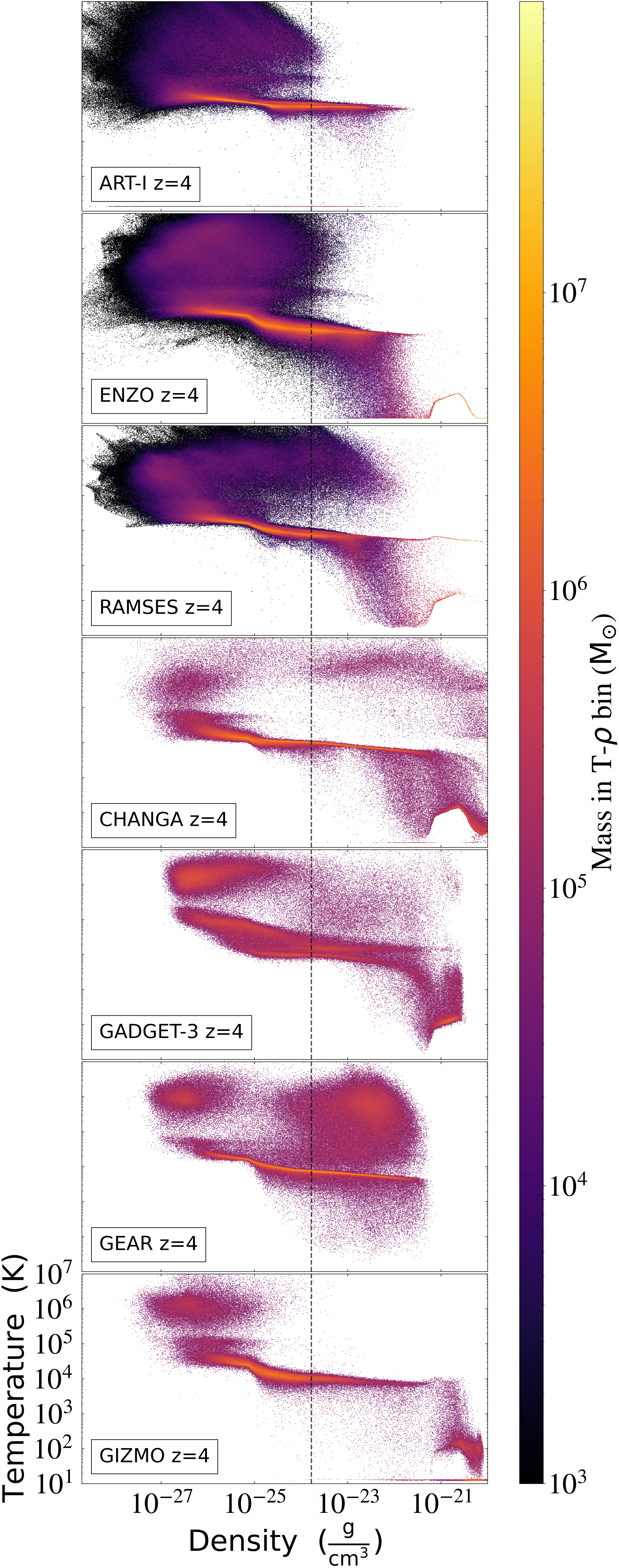}
        \caption{The $z=4$ composite of 2-dimensional PDF of density and temperature for the gas within the mean $R_{200}$ among the codes  ($\sim 25.4$ kpc)  from the target galaxy's center in the {\tt CosmoRun} simulations.
        It is similar to Figures \ref{Fig01}, \ref{Fig1} and \ref{Fig3}; but, unlike the previous figures, a sphere of  $R_{200}$ encloses the main galaxy and CGM, but not the IGM.  
        Colors represent the total gas mass in each 2-dimensional bin. 
        A black dashed vertical line marks the density threshold for star formation. 
        See Section \ref{sec:results} for more information on {\tt CosmoRun}, and Section \ref{sec:CosmoRunGas} in particular on this figure.
        }
        \label{Fig13}
\end{figure}

\begin{figure*}
        \centering
        \includegraphics[scale=0.68]{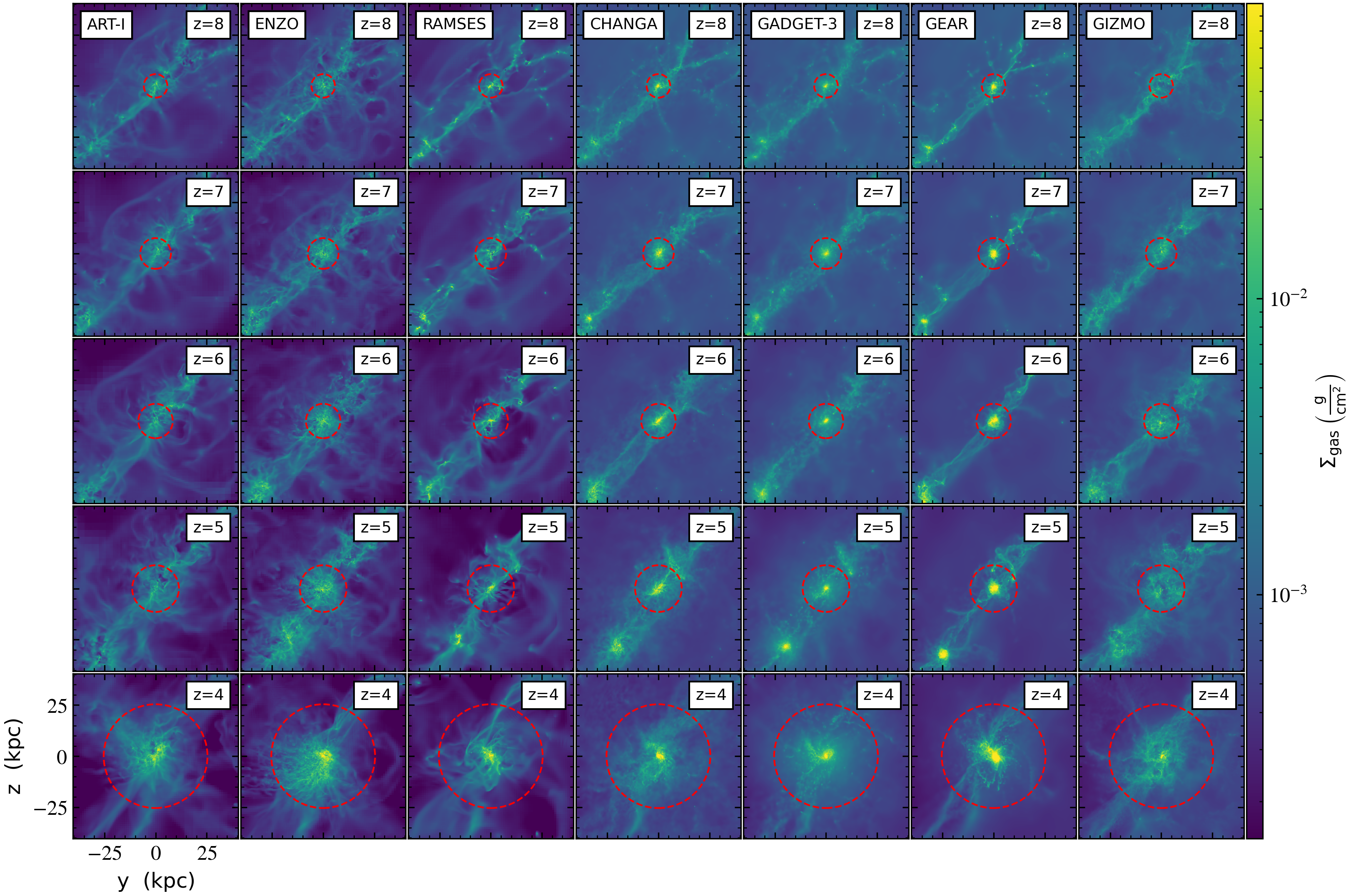}
        \caption{Gas surface densities at $z=8$ to 4 from our final {\tt CosmoRun} simulation suite, centered on the center of mass of stars and dark matter belonging to the target galaxy progenitor.
        Here and in the following figures we indicate the mean $R_{200}$ among the codes at each redshift with a red dashed circle (5.8, 7.5, 8.4, 11.4 and 25.4 proper kpc at $z=8, \, 7, \, 6, \, 5$ and 4, respectively). 
        Units are proper kpc. 
        The projections along the other axes are available as digital supplements to this article.  
        See Section \ref{sec:CosmoRunGas} for more information on {\tt CosmoRun} and this figure.}
        \label{Fig15}
        \vspace{2mm}
\end{figure*}

\begin{figure*}
        \centering
        \includegraphics[scale=0.68]{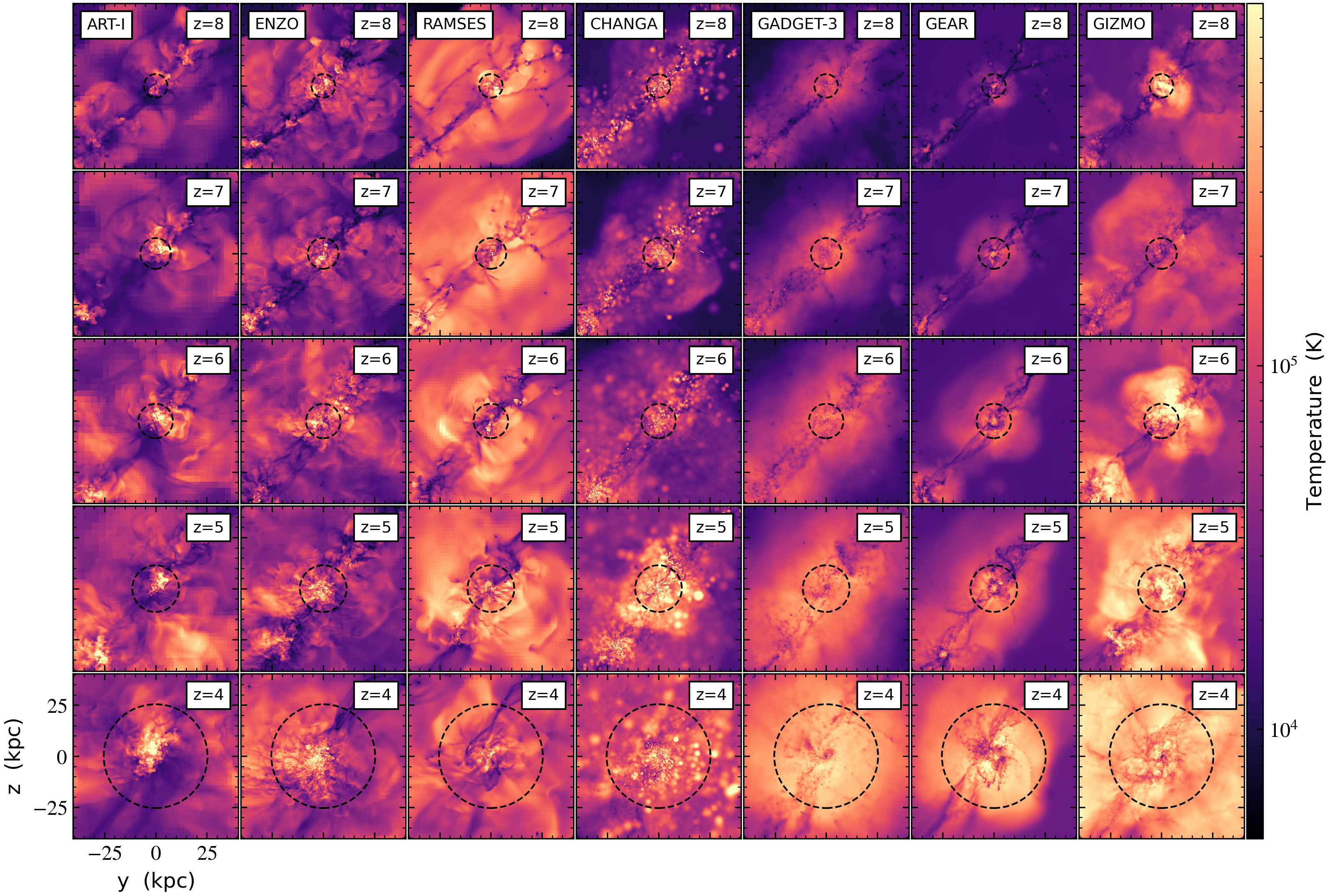}        
        \caption{Similar to Figure~\ref{Fig15}, but now showing density-square-weighted projections of gas temperature  in our {\tt CosmoRun} simulation suite.
        Units are proper kpc.
        See Section \ref{sec:CosmoRunGas} for more information on {\tt CosmoRun} and this figure.}
        \label{Fig14}
        \vspace{2mm}
\end{figure*}

\vspace{1mm}

\subsection{Gas Properties}\label{sec:CosmoRunGas}

Because deviations in stellar feedback are better reflected in gas, gas properties in simulations can be used to compare and calibrate the stellar feedback prescriptions employed. 
It is not in the scope of this paper to determine which stellar feedback in which code better fits the observations. 
Instead, we aim to show which gas properties are more sensitive to feedback, and to provide the community with a common ground to make new comparisons. 
In this subsection, we present only a general analysis of the gas properties. 
This first analysis is currently being extended and will be presented in a future paper focused on the evolution of the CGM.

\subsubsection{Gas Density and Temperature}

\begin{figure}
        \centering
        \vspace{2mm}
        \includegraphics[scale=0.38]{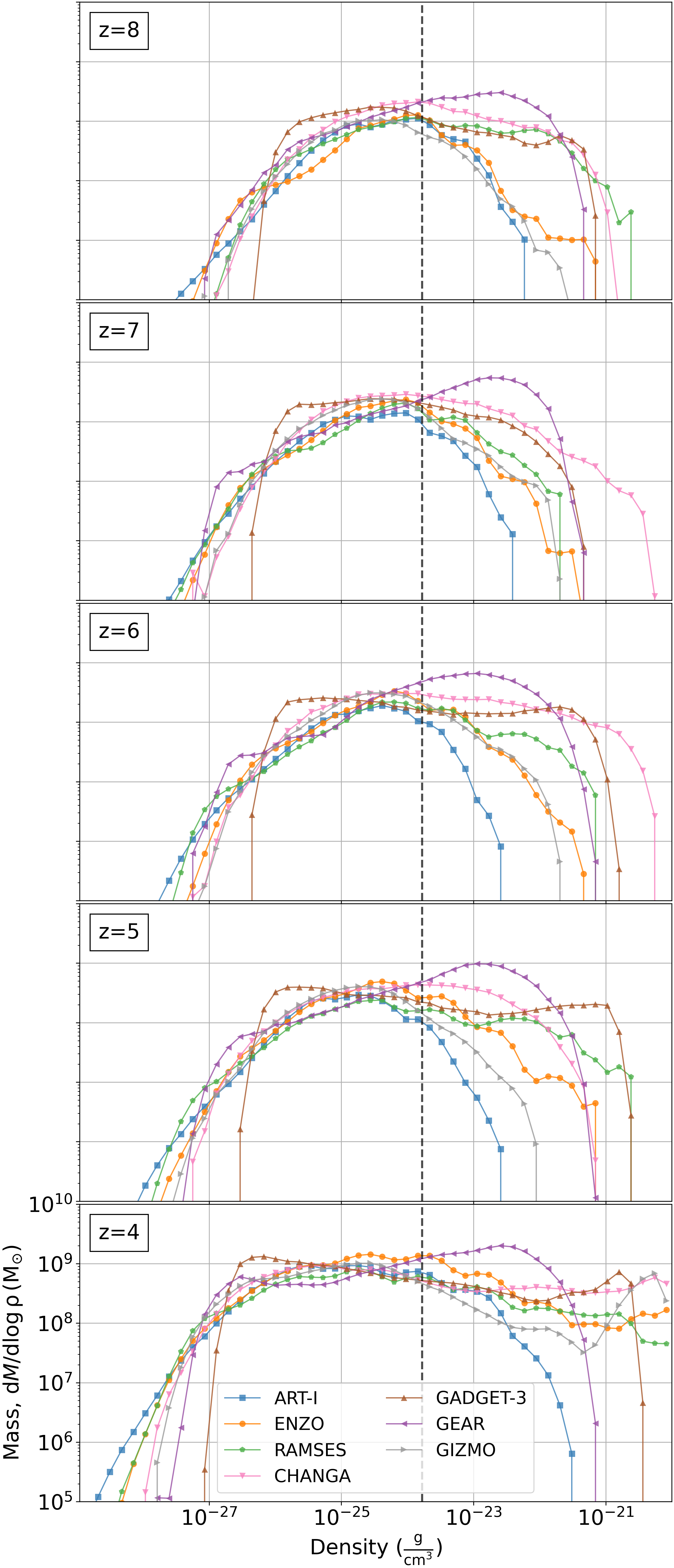}
        \caption{Distribution of gas mass as a function of gas density at $z=8,\,7,\,6,\,5$ and $4$ from our {\tt CosmoRun} simulation suite.
        Each panel is for all the gas inside the target progenitor's $R_{200}$. 
        The vertical black dashed line denotes the star formation threshold, $n_{\rm H, \,thres} = 1\, {\rm cm}^{-3}$.  
        See Section \ref{sec:CosmoRunGas} for more information on {\tt CosmoRun} and this figure.}
        \label{Fig11}
\end{figure}

The first figure of this subsection, Figure~\ref{Fig13}, displays the gas density-temperature PDF, that can be compared with Figures \ref{Fig01}, \ref{Fig1} and \ref{Fig3} from our calibration steps {\tt Cal-1} to {\tt Cal-3} (see Section \ref{sec:calibration}). 
Note that, in this plot, we only show the gas inside $R_{\rm 200}$ (see the caption of Table~\ref{tab:Generalparams}), while Figures \ref{Fig01}, \ref{Fig1} and \ref{Fig3} include gas out to the IGM. 
From Figure~\ref{Fig13}, we see that, once the stellar feedback is activated, the convergence we always get is only the shape of the $\sim10^4$ K cooling curve. 
Notable differences between the codes in Figure~\ref{Fig13} include:
{\it (1)} The blue bins with small mass per bin  in the mesh-based codes reflecting very diffuse gas, that are not well represented in the particle-based codes (as discussed in  Sections~\ref{sec:Cal2_comp} and \ref{sec:results_cal3}). 
{\it (2)} The total gas mass $M_{\rm gas}$ inside $R_{\rm 200}$  changes significantly between codes due to the different stellar feedback strategies adopted (see Section \ref{code-phy}) and the ``timing discrepancies'' (see Section \ref{sec:results_cal4} and footnote \ref{footnote:timing}), for which a clear example appears when comparing the total $M_{\rm gas}$ (number of bins and colors) in $R_{\rm 200}$ of e.g., {\sc Art-I} and {\sc Changa} (see also Figure \ref{Fig11}). 
The exact timing of a major merger occurring at around $z\sim4$ partly explains the discrepancy in the PDF between different codes.  
For example, while {\sc Art-I} still undergoes the merger at $z=4$, other codes already experienced it at slightly earlier times (see Section \ref{sec:CosmoRunStars} and Figure \ref{Fig24}).
{\it (3)} In addition to driving the gas out of $R_{\rm 200}$, the different stellar feedback strategies may also instigate other differences in the PDF, in particular in the warm-hot gas phase ($\sim 10^{5-7}$ K) above the threshold for star formation, $n_{\rm H, \,thres} = 1\, {\rm cm}^{-3}$.  
Indeed, the gas in star-forming regions is sensitive to variations in the stellar feedback strategies used to release energy and momentum from newly-formed stars. 
Particularly, the use of a delayed cooling strategy (in {\sc Ramses}, {\sc Gadget-3} and {\sc Gear}) may result in the accumulation of warm-hot gas in a dense state, around star forming regions.
The superbubble feedback scheme used in {\sc Changa} produces a similar effect on the warm-hot dense gas.
{\it (4)} Lastly, the cold diffuse gas near the CMB floor, visible only in {\sc Art-I}, is due to the code's stochastic star formation recipe and its particular pressure floor implementation (as discussed in footnote \ref{footnote:art-i-stochastic} and Section~\ref{sec:results_cal3}).\footnote{As a final note to Figure \ref{Fig13}, the gas at $\gtrsim 10^{-21}\,{\rm g\,cm^{-3}}$ is seen heated up to $\sim 10^2$ K (except in {\sc Art-I} and {\sc Gear} in which such dense gas is nonexistent for the moment). 
This heated gas is caused by {\sc Grackle}'s redshift-dependent UVB with self-shielding (Section~\ref{common-phy}), and is  observed even in a simple one-zone test using {\sc Grackle}. 
The source of the heating is assumed to be re-emission of absorbed radiation inside the dense gas cloud. 
The shielded {\sc Cloudy} tables were made by integrating into the star-forming cloud for a distance set by the Jeans length at a given density and temperature  (with a maximum of 0.1 kpc). 
Over this length, UVB radiation absorbed by the outer layers of the cloud can be re-emitted, causing some heating on the inner layers.
We caution {\sc Grackle} users when they use the default shielded {\sc Cloudy} table provided with the package  (e.g., depending on the simulation setup and resolution, one may want to disable UVB above a certain density).}    

To better illustrate the effect of stellar feedback on the gas in the galaxy, the CGM, and the IGM, we show the evolution of the projected density and temperature in each code in Figures~\ref{Fig15} and \ref{Fig14}. 
The mean virial radius, $R_{\rm 200}$, at each redshift (see the Figure~\ref{Fig15} caption) is marked with a red/black dashed circle.
In these figures, we confirm the differences in the spatial distribution and thermal structure of gas, due to variations in the stellar feedback strategies, despite the fact that all the participating codes produce similar stellar mass at our target epoch, $z=4$.
Although differences in gas density and  temperature may appear dramatic in Figures~\ref{Fig13} to \ref{Fig14}, we find a good agreement in the density distribution, especially in the nonextreme density range. 
This result can be observed in Figure~\ref{Fig11}, where we show the evolution of the gas density PDF of all the gas inside $R_{\rm 200}$ from $z=8$ to $z=4$. 
We clearly see that most codes agree on the total gas mass --- the area below the curve --- in the intermediate density range, [$\sim10^{-27}$, $\sim10^{-23}$] ${\rm g\,cm^{-3}}$. 
Obviously, discrepancies in the lowest and highest density bins exist, produced by various reasons discussed in Figure \ref{Fig13} (note that Figure \ref{Fig11} shows the values of Figure \ref{Fig13} integrated along its $y$-axis).  

\vspace{3mm}

\subsubsection{Gas Metallicity}\label{gas-metallicity}
    
Metallicity is a good tracer of changes in galactic evolution. 
The metal content of gas inside the galaxy and its CGM, depends on how efficiently the outflows remove the metal-rich gas from the dense star-forming regions. 
The metal enrichment of the IGM is also dictated by the ouflows, as the IGM is the recipient of the gas pushed out of the virial radius. 
The exchange of metals between the CGM and IGM also determines the gas evolution in time on the density-temperature plane, as it strongly affects how quickly the gas cools and regulates the interplay between star formation and feedback.  
Metallicity indeed provides important information on the differences between the feedback schemes employed, and their ability to fit observations \citep{2015MNRAS.448..895S,2019ApJ...886...91K,2020ApJ...900....9L}. 

Before presenting the next figures on metallicities, it is important to remind the readers that all code groups used metal yields in supernovae that are similar to the ones in the {\it AGORA} common physics (see Section~\ref{common-phy}). 
Using metal yields similar to the common ones allows us to conjecture that the differences observed in gas metallicity are explained mostly by the variations in stellar feedback --- and/or the metal diffusion schemes --- presented in Section~\ref{code-phy}. 
As a consistency check, in each {\tt CosmoRun} simulation we have computed the ratio of the total metal mass and the total stellar mass  inside the entire simulation box at $z=4$ (i.e., ``effective'' metal yields in the fourth column of Table \ref{tab:feedback}).    
Our calculation confirms that, although each code group is using its favorite metal production strategy, its ``effective'' yield value matches what each group assumes in the code's deposit scheme, and is in agreement within less than half a dex from other codes.

\begin{figure*}
        \centering
        \includegraphics[scale=0.68]{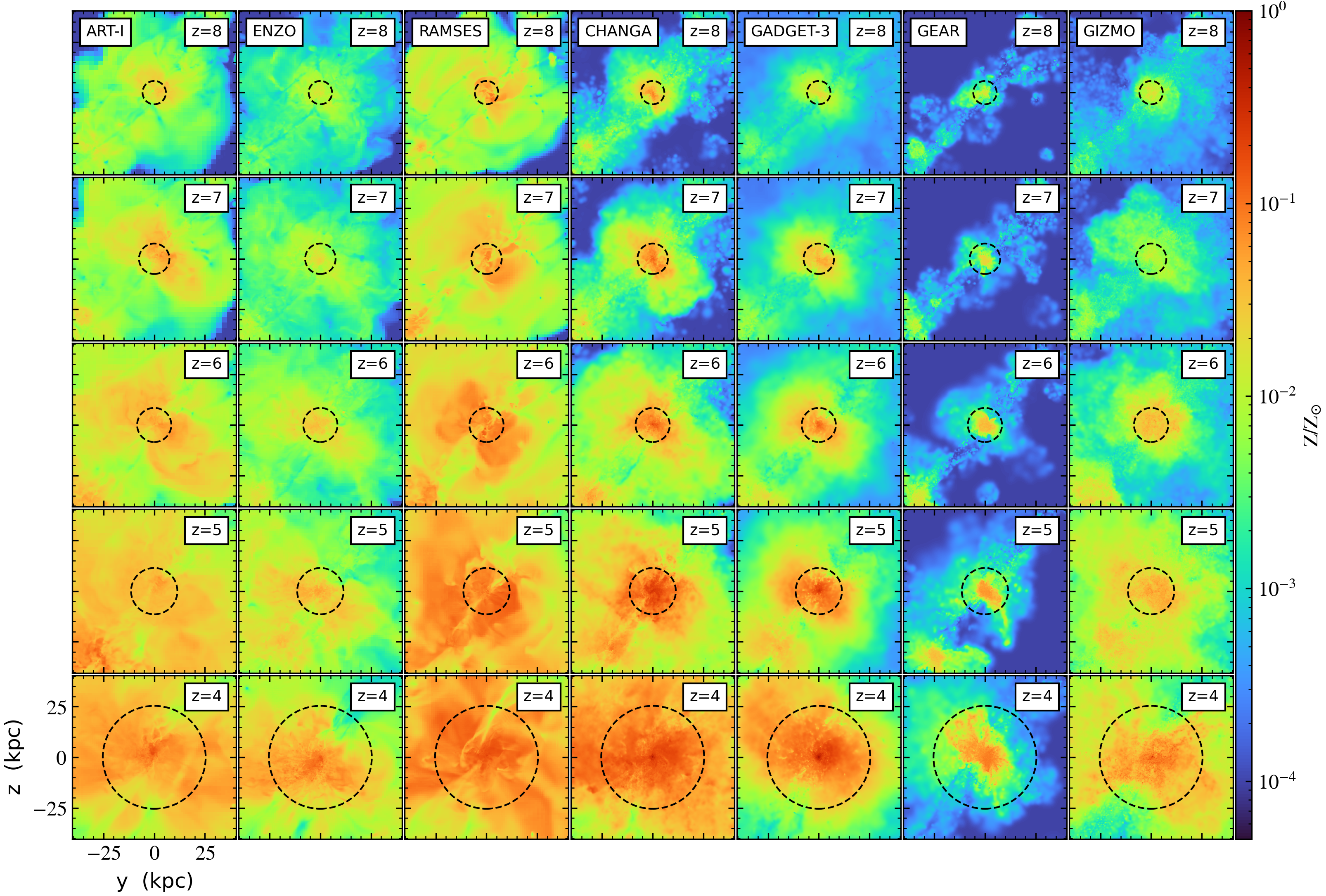}
        \caption{Similar to Figure~\ref{Fig15} and \ref{Fig14}, but now showing density-square-weighted projections of gas metallicity in our {\tt CosmoRun} simulation suite.
        Colors represent the metallicity in units of Z$_{\odot}$. 
        Units are proper kpc.
        See Section \ref{sec:CosmoRunGas} for more information on {\tt CosmoRun} and this figure.}
        \label{Fig19}
        \vspace{2mm}        
\end{figure*}

First, in Figure~\ref{Fig19}, we show the projected gas metallicity at $z=8,\,7,\,6,\,5$ and $4$. 
It is important to mention that a correct interpretation of this figure requires the information on the total gas distribution (Figure~\ref{Fig15}), e.g., most metals in {\sc Gear} are in low-metallicity dense gas in the inner parts of the halo.
Some codes show high metallicity around the main galaxy (e.g., {\sc Ramses}, {\sc Changa}, and {\sc Gadget-3}), while others exhibit lower values (e.g., {\sc Art-I}, {\sc Enzo}, and {\sc Gizmo}). 
The former codes are the ones that tend to keep gas and metals around the star-forming regions, while the latter codes are able to push them out to the CGM, or even the IGM (see also Figure~\ref{Fig30}). 
The discrepancy seen here is also because the spatial distribution of metals is highly sensitive to how efficient the stellar feedback is at driving the metal-enriched outflows (see Figure~\ref{Fig30}), and to how efficient the metal diffusion is at polluting the neighboring cells/particles.
    
We reach a similar conclusion by analyzing the probability distribution function of metallicity and metal mass in Figures~\ref{Fig23} and \ref{Fig20}, respectively.  
Here we include only the gas inside a sphere of $R_{\rm 200}$ from the target progenitor's center. 
Figure~\ref{Fig23} shows that {\sc Ramses}, {\sc Changa}, {\sc Gadget-3} and {\sc Gizmo} exhibit large amounts of high-metallicity  ($\gtrsim 1\,\zsun$) gas in and around the main galaxy, while {\sc Art-I}, {\sc Enzo}, and {\sc Gear} shows less amounts.
This difference confirms that the overall gas metallicity distribution depends strongly on the efficiency of stellar feedback.   
Furthermore, in Figure~\ref{Fig20}, --- while the global features in the PDF have been discussed in the section relevant to Figure~\ref{Fig13} --- we find variations in the total metal mass kept inside $R_{\rm 200}$. 
The stellar feedback in {\sc Art-I} and {\sc Enzo} rapidly push the metals out to the low-density and low-metallicity gas in the CGM and then to the IGM, leaving only a few dense star-forming regions with high metallicity. 
In contrast, the remaining codes keep most of the metals inside $R_{\rm 200}$, showing more regions with high metallicity in the gas density-temperature plane, particularly inside the regions of delayed cooling.

\vspace{5mm}

\subsection{Stellar Properties} \label{sec:CosmoRunStars}

In this section, we carry out a global analysis of the stellar components in the {\tt CosmoRun} simulations, but only focusing on their spatial distribution and metallicity. 
A more detailed analysis of the stellar component, including kinematics, SFHs, in-situ vs. ex-situ origin, and low-$z$ evolution will be presented in a future paper by the {\it AGORA} Collaboration.  

In Section~\ref{sec:results_cal4} for {\tt Cal-4}, we have examined the stellar mass growth histories (Figure~\ref{Fig6}). 
There, we detect occasional increases in stellar masses in most codes --- the kinds of increases that are not contemporaneous between the codes. 
In fact, these are signs of the major mergers, which can be best observed in the stellar surface density maps  in Figure~\ref{Fig24}. 
The mean virial radius, $R_{\rm 200}$, at each redshift (see the Figure~\ref{Fig15} caption)  is marked with a white dashed circle in each panel.
In this figure, it is easier to perceive that major/minor mergers do not occur at the same time in every simulation due to the aforementioned ``timing discrepancy'' (see Sections \ref{sec:results_cal4} and \ref{sec:CosmoRunGas}). 
The $z=4$ row is particularly interesting. 
By $z=4$, most codes have gone through a recent major merger event, but they are at different stages of halo relaxation. 
This observation warns us of the need to be careful when comparing properties of galaxy-scale systems in cosmological simulations between different codes; it is indeed prudent to avoid the times when a strong perturbation is ongoing. 
Simulations presented here will be further analyzed in a future paper, also at lower redshifts when major mergers are rare and comparisons are more straightforward.

\begin{figure}
        \centering
        \vspace{3mm}        
        \includegraphics[scale=0.48]{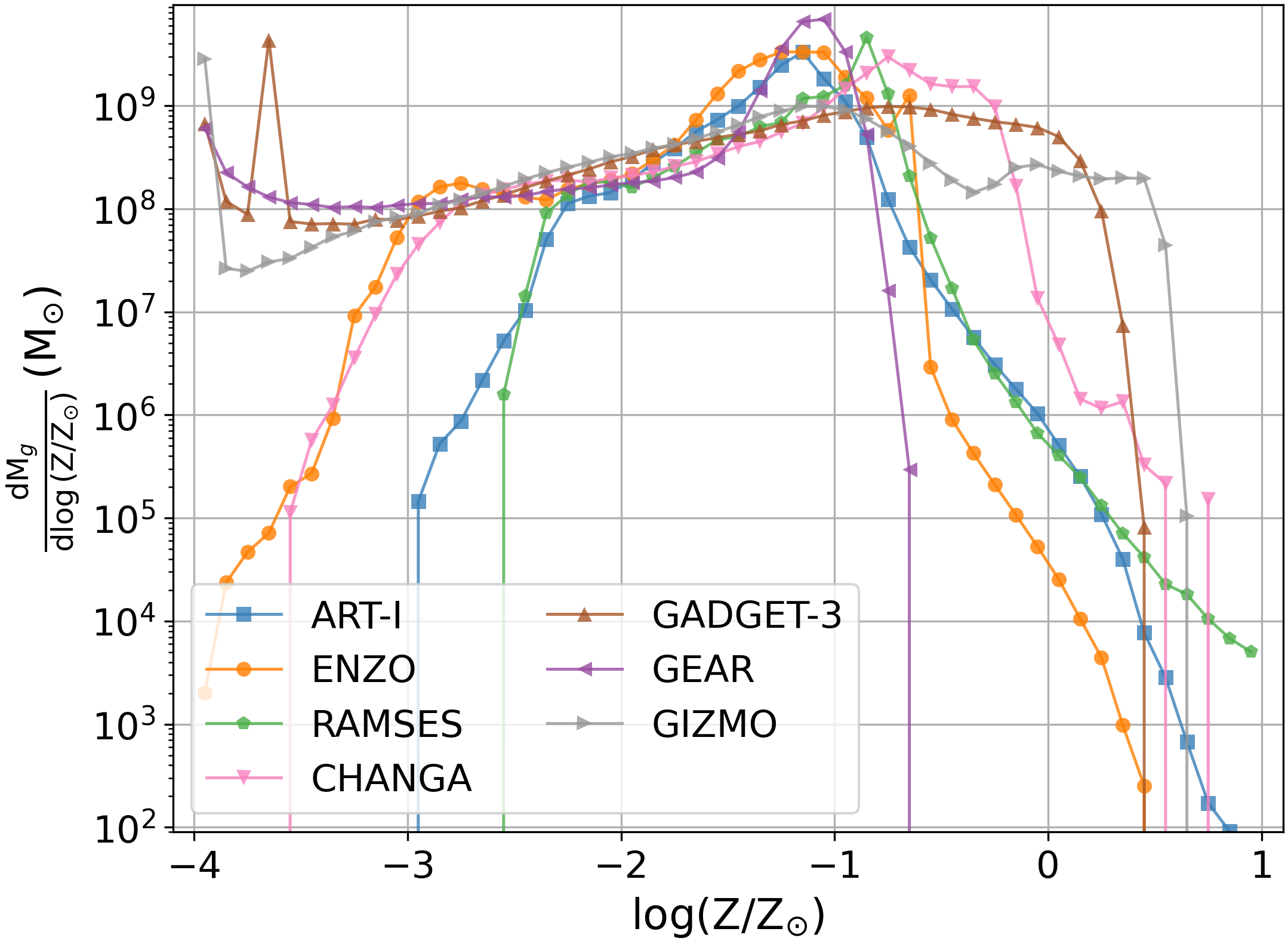}
        \caption{Distribution of gas mass as a function of gas metallicity at $z=4$ for all the gas inside the target progenitor's $R_{200}$ in our {\tt CosmoRun} simulation suite.
        The y-axis range is kept identical as in Figure~\ref{Fig25} for easier comparison.
        See Section \ref{sec:CosmoRunGas} for more information on {\tt CosmoRun} and this figure.}
        \label{Fig23}
        \vspace{0mm}         
\end{figure}

\begin{figure}
        \centering
        \vspace{2mm}         
        \includegraphics[scale=0.25]{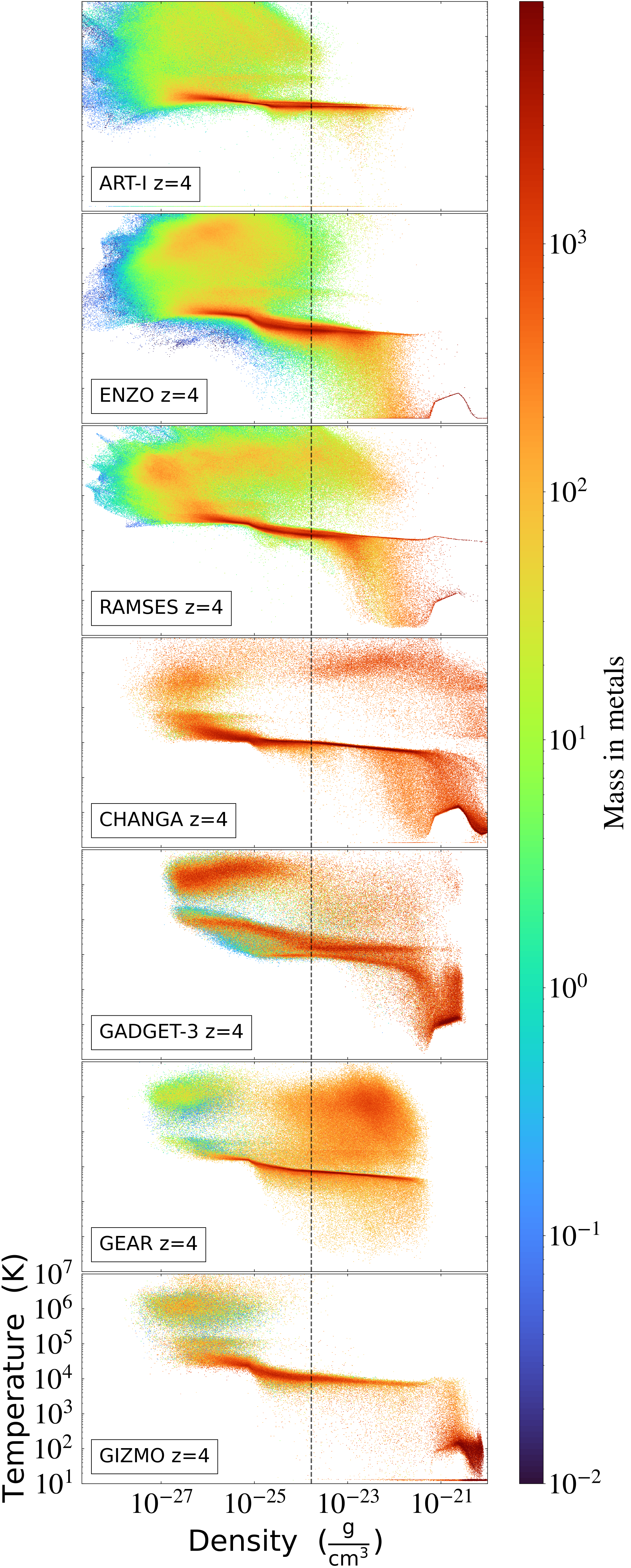}        
        \caption{Similar to Figure~\ref{Fig13}, but now with colors representing the total metal mass in each 2-dimensional bin in our {\tt CosmoRun} simulation suite.   
        Note that the PDF is for the gas within $R_{200}$ from the center of the target galaxy in the {\tt CosmoRun} simulations.  
        A sphere of radius $R_{200}$ encloses the main galaxy and CGM, but not the IGM.  
        See Section \ref{sec:CosmoRunGas} for more information on  {\tt CosmoRun} and this figure.
        }
        \label{Fig20}
\end{figure}
    
\begin{figure*}
        \centering
        \includegraphics[scale=0.68]{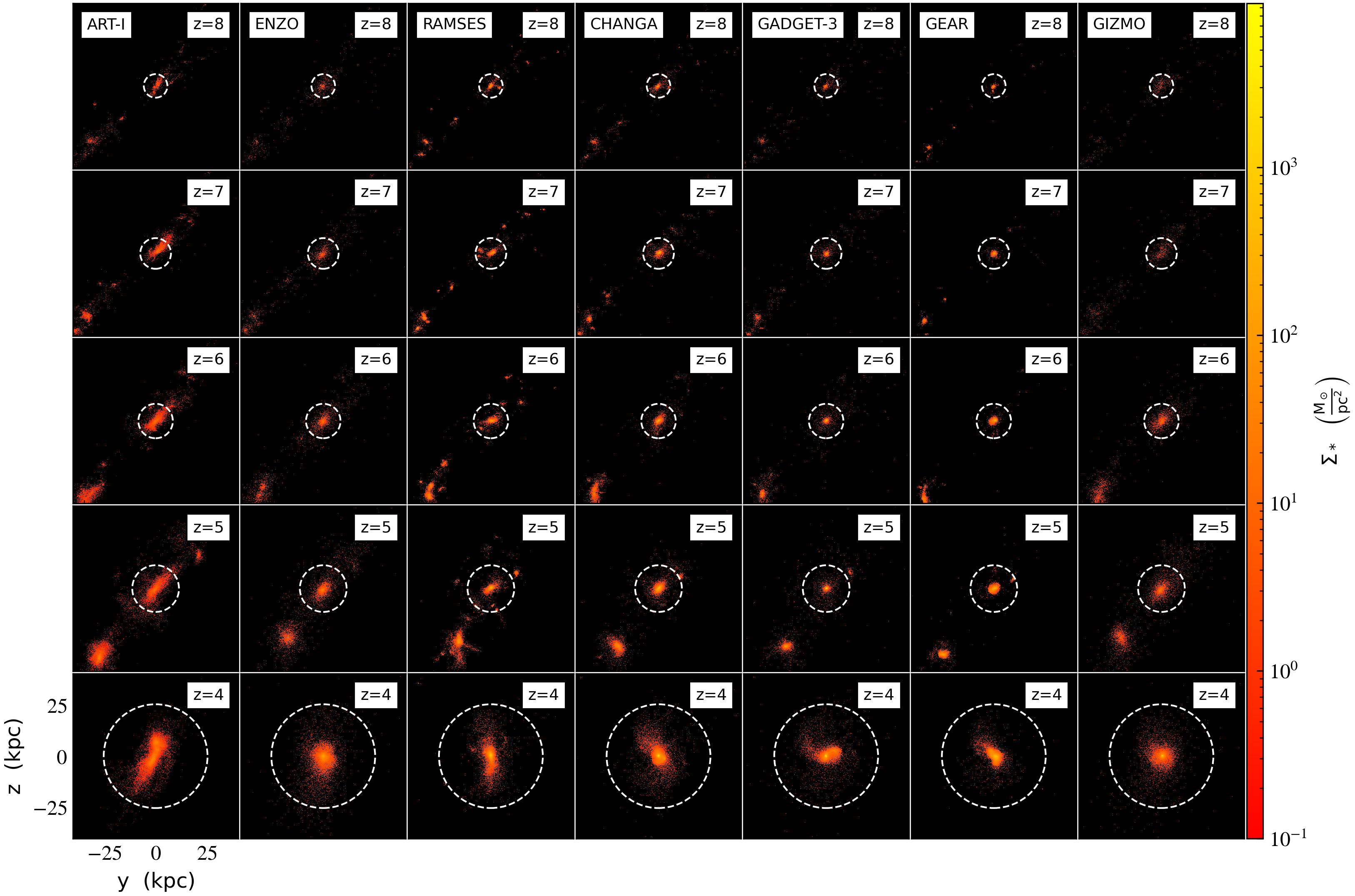}
        \caption{Similar to Figure~\ref{Fig15}, \ref{Fig14} and \ref{Fig19}, but now presenting stellar surface densities  from our {\tt CosmoRun} simulation suite.
        Colors represent the total stellar mass in each 2-dimensional bin.
        Units are proper kpc.
        See Section \ref{sec:CosmoRunStars} for more information on {\tt CosmoRun} and this figure.}
        \label{Fig24} 
        \vspace{2mm}        
\end{figure*}  

\begin{figure}
        \centering
        \vspace{2mm}
        \includegraphics[scale=0.48]{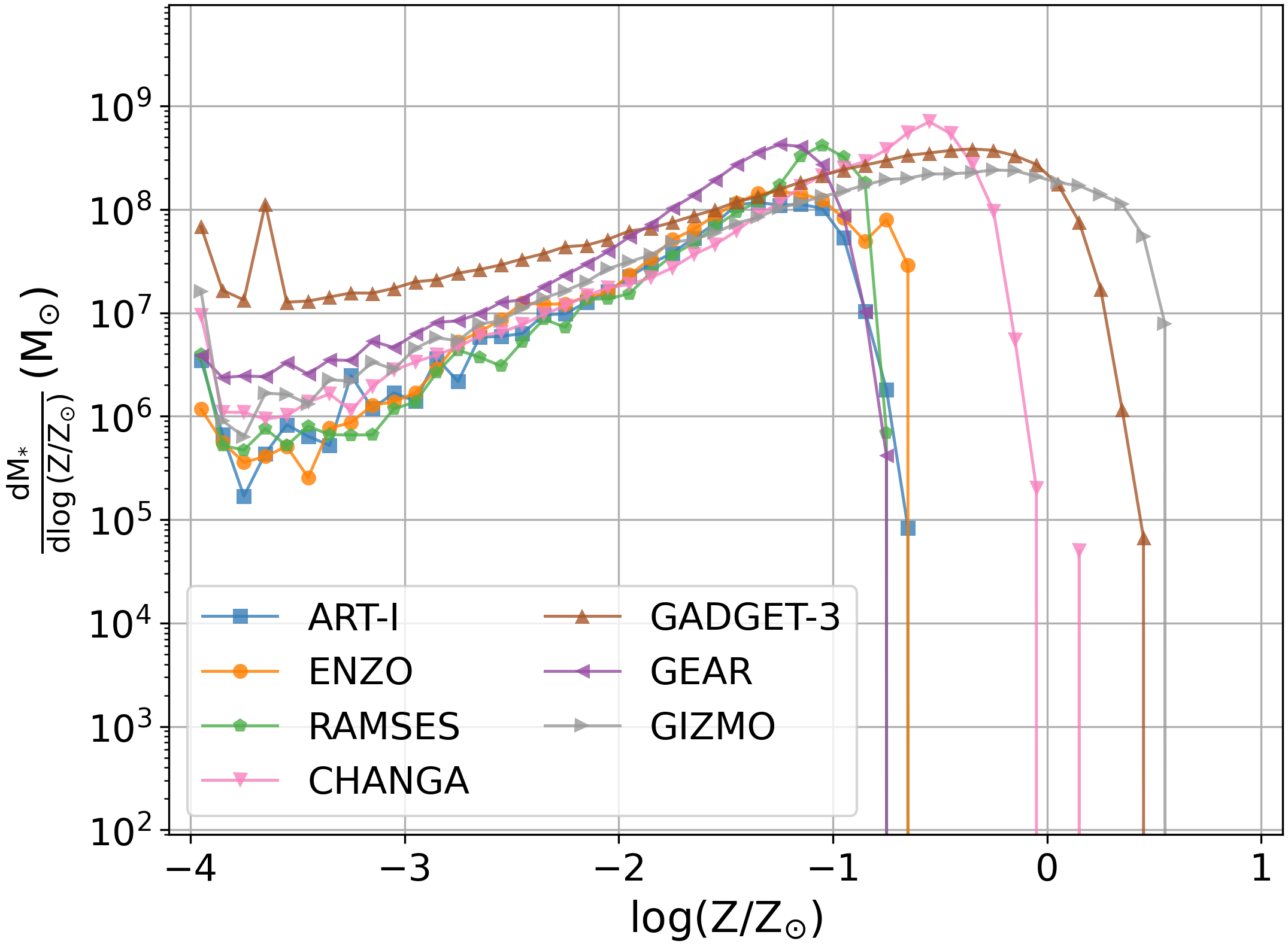} 
        \caption{Distribution of stellar mass as a function of stellar metallicity at $z=4$ for all the stars inside the target progenitor's $R_{200}$ in our {\tt CosmoRun} simulation suite. The y-axis range is kept identical as in Figure~\ref{Fig23} for easier comparison.
        See Section \ref{sec:CosmoRunStars} for more information on {\tt CosmoRun} and this figure.}  
        \label{Fig25}
        \vspace{2mm}        
\end{figure}

We conclude this subsection by investigating stellar metallicities and comparing the results with the distribution of metals in the gas component. 
By construction, stars  form in regions where gas reaches the imposed star formation threshold, thus they inherit the properties of their progenitor gas. 
Among the inherited properties, metallicity is the one that should follow a similar trend between stars and the high-density gas. 
Additionally, in the gas metallicity PDF within $R_{\rm 200}$ (Figures~\ref{Fig23} and \ref{Fig20}), we expect to find that a significant fraction of gas in the high-density, high-metallicity bins is star-forming. 
This argument is in agreement with what we observe in Figure~\ref{Fig25}, in which we show the stellar mass per metallicity bin. 
As can be also inferred from Figures~\ref{Fig11} and \ref{Fig23}, the stellar metallicity distribution peaks at a similar value to the gas metallicity in each code.
Nevertheless, the distribution tends to be narrower in the stellar metallicities (Figure~\ref{Fig25}) than in the gas metallicities (Figure~\ref{Fig23}), as most star particles form in the densest pockets of gas.  
The low-metallicity stars could be either the early generation of stars formed in the gas that has not been heavily metal-enriched yet, or the later generation of stars formed in the CGM only lightly metal-enriched by galactic outflows.  

\vspace{1mm}

\subsection{Circumgalactic Medium (CGM) Properties}\label{sec:CosmoRunCGM}

The {\it AGORA} collaboration plans to  work on a full analysis of the CGM properties and evolution of the presented {\tt CosmoRun} simulations, from high $z$'s down to $z=2$. 
The results of this extensive analysis will be presented in a forthcoming paper. 
In this section, however, we demonstrate how the multi-platform studies like {\it AGORA} could be useful to better understand the thermal and kinematic states of the CGM, in which disparities exist between contemporary cosmological simulations carried out with different codes, by presenting the first analysis of gas kinematics in four different temperature bins at $z=4$. 
The temperature bins are defined following the observationally-motivated temperature thresholds proposed in \citet{RocaFabrega19} and in \citet{2021MNRAS.501.4948S}.

\begin{figure*}
        \centering
        \vspace{2mm}
        \includegraphics[scale=0.4]{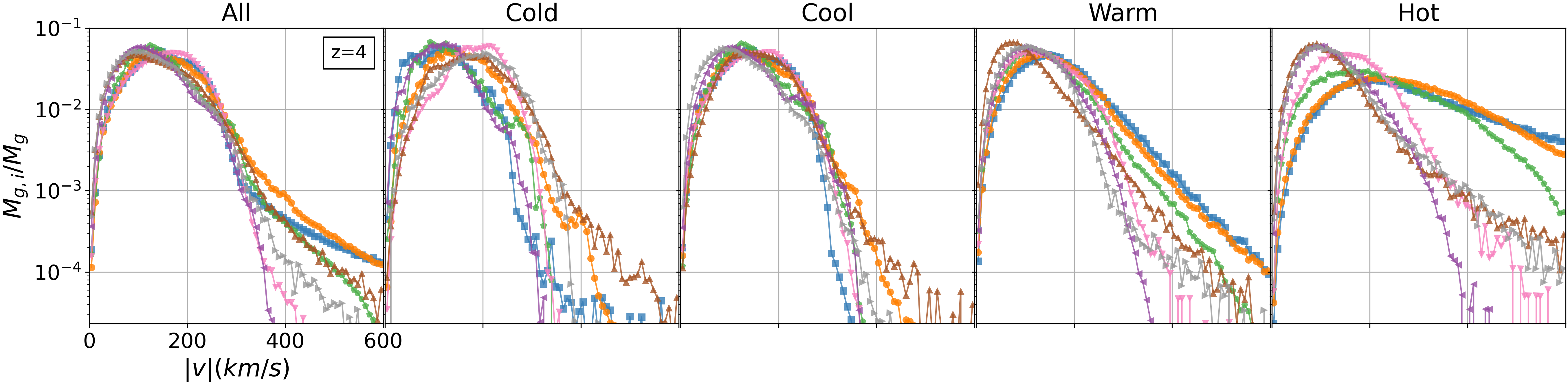} 
        \includegraphics[scale=0.4]{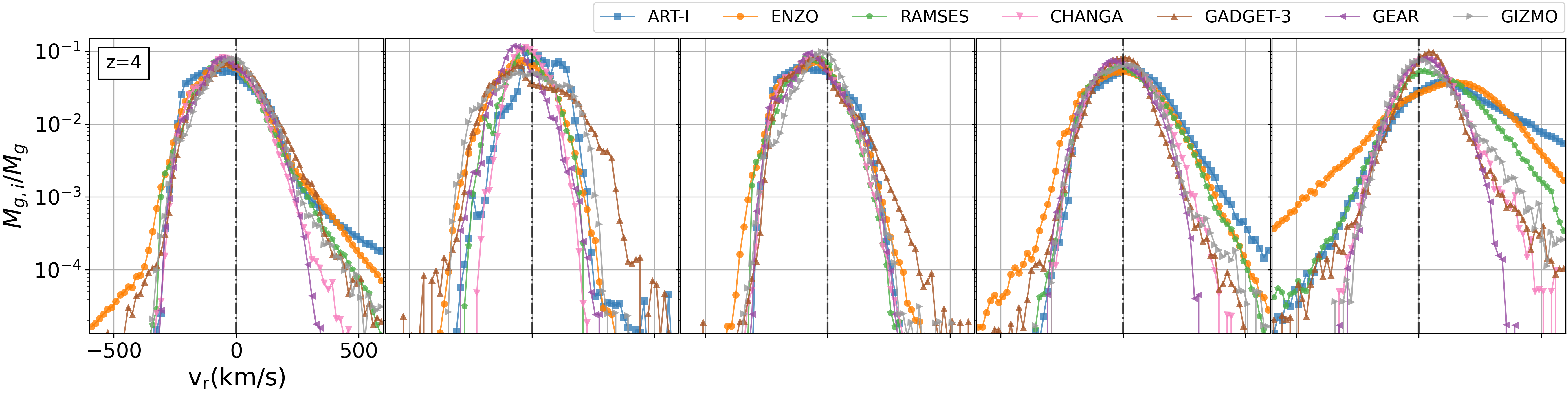}        
        \caption{Distribution of gas mass as a function of velocity at $z=4$ --- velocity magnitude ({\it top}) and radial velocity ({\it bottom}) --- for the gas inside the target progenitor's $R_{200}$ in our {\tt CosmoRun} simulation suite.
        The $y$-axis indicates the fraction of gas mass in each velocity bin with respect to the total mass in each temperature phase.   
        The panels are for all the gas, cold gas ($T<10^{3.8}\,{\rm K}$), cool gas ($10^{3.8}<T<10^{4.5}\,{\rm K}$), warm gas ($10^{4.5}<T<10^{6.5}\,{\rm K}$), and hot gas ($T>10^{6.5}\,{\rm K}$) from left to right. 
        See Section \ref{sec:CosmoRunCGM} for more information on {\tt CosmoRun} and this figure.}   
        \label{Fig30}
        \vspace{3mm}
\end{figure*}

In Figure~\ref{Fig30}, we show the probability distributions of the velocity magnitude (top row) and the radial velocity (bottom row) for the gas  inside a sphere of radius $R_{\rm 200}$ from the center of the target progenitor galaxy.
The panels are for all the gas, cold gas ($T<10^{3.8}\,{\rm K}$), cool gas ($10^{3.8}<T<10^{4.5}\,{\rm K}$), warm gas ($10^{4.5}<T<10^{6.5}\,{\rm K}$), and hot gas ($T>10^{6.5}\,{\rm K}$) from left to right. 
The velocity magnitude PDFs (top row) show that there is a reasonably good agreement on the kinematics of the gas. 
This agreement is particularly good in the cool and warm gas; in these temperature phases, the mesh-based codes and the particle-based codes agree well with each other. 
The convergence is not as good in the hot gas, though, where {\sc Art-I} and {\sc Enzo} exhibit slightly larger gas fraction with high velocity than the rest of the participating codes, due to stronger feedback-driven outflows (rightmost panel; as discussed in Section \ref{sec:CosmoRunGas}).  
The {\sc Ramses} run presented here shows lower velocities than {\sc Art-I} and {\sc Enzo} in the hot gas component as expected from our analysis of metal distribution (see a full discussion in Section~\ref{sec:CosmoRunGas}). 
Additionally, in the {\sc Changa}, {\sc Gadget-3}, {\sc Gear} and {\sc Gizmo} runs, the hot gas with the largest velocities typically belongs to regions with very low density that are not well represented by their particle-based approach. 
In agreement with our conclusions on the gas metallicity distribution (see Section \ref{gas-metallicity}), {\sc Gear} generates the slowest outflows, keeping most of the metals in the dense gas around the galaxy.

In the bottom row of Figure~\ref{Fig30}, we show the distribution of gas mass in radial velocity bins. 
Radial velocity informs us of the presence of inflowing or outflowing gas, and the strength thereof. 
As discussed in the previous paragraph and in Section \ref{sec:CosmoRunGas}, the strong feedback-driven outflows in {\sc Art-I} and {\sc Enzo} are evident in the hot gas phase (rightmost panel; also in the warm phase for  {\sc Art-I}).  
This outflowing hot gas transports a large fraction of metals to the IGM, leaving the CGM in {\sc Art-I} and {\sc Enzo} with lower metallicity relative to the other codes. 
The {\sc Ramses}, {\sc Changa}, {\sc Gadget-3}, {\sc Gizmo} and particularly {\sc Gear} runs do not show as strong outflows as in {\sc Art-I} or {\sc Enzo}, keeping most of the metals and gas inside the CGM (as also seen in Figures \ref{Fig11} and \ref{Fig19}). 
The cool gas follows a smooth distribution centered at zero velocity but slightly inflowing (3rd panel from the left), with a very good agreement among all the codes. 

The very preliminary analysis of the gas properties in the CGM and, in particular, of its kinematics in four different temperature bins, teaches us that the kinematics of the cold and hot gas is a good tracer of differences in the adopted stellar feedback prescriptions. 
We suggest that the research groups interested in testing their feedback models include the study of cold and hot gas kinematics in their comparisons.

\vspace{1mm}

\section{Discussion and Conclusion} \label{sec:conclusion}

\vspace{1mm}

In this paper, we have presented a suite of seven high-resolution cosmological zoom-in simulations to $z=4$ of a halo with a Milky Way mass at $z=0$, obtained using seven contemporary astrophysical simulation codes --- 3 AMR codes and 4 SPH codes --- widely used in numerical galaxy formation. 
The physics prescriptions in the simulations include the common gas cooling and heating by {\sc Grackle-v3.1.1} that are similar to what was used in the previous {\it AGORA} comparisons, and the standardized {\it AGORA} subgrid physics such as star formation and stellar evolution (Section \ref{common-phy}). 
However, the code groups participating in the comparison  use the stellar feedback prescription that resembles the most widely used in their code community for research (Section \ref{code-phy}).
The simulations also account for the effects of cosmological processes such as the expansion of the Universe, and the cosmic UVB radiation emitted by massive stars and quasars. 

The simulations presented here have been obtained after a careful, four-step process of calibrations (Section \ref{sec:calibration}). 
The calibration strategy designed by the Collaboration is to reduce the number of tunable simulation parameters to be accounted for when studying the effects of stellar feedback on galaxy evolution. 
By completing this set of  calibrations, the participating code groups establish a common ground to make a robust and unbiased comparison of different simulations focusing on stellar feedback effects on the gas and SFH of the target galaxy.
The calibration procedure includes four steps. 
In the first step ({\tt Cal-1}) the code groups control the effects of the different gravity and hydrodynamics solvers, and refinement strategies in radiative cooling/heating-free simulations. 
In the second step ({\tt Cal-2}), we ensure that the {\sc Grackle} cooling and UVB are correctly implemented in each code.
The third step ({\tt Cal-3}) aims for convergence in the total stellar mass produced with the common star formation prescription in stellar feedback-free simulations. 
Finally, in the last calibration step ({\tt Cal-4}), we ask each code group to test a stellar feedback prescription that is as close to the most commonly used one in each code community as possible, while aiming for convergence in the stellar-to-halo mass ratio at $z=4$  to the prediction by semi-empirical models.
Designing and executing the calibration procedure has required formidable efforts by the Collaboration members to (re)run the simulations while revising, when necessary, the physical prescriptions they use for the final cosmological simulations. 

After all the participating code groups successfully completed  the calibration steps, we reach a suite of cosmological zoom-in simulations with very similar mass assembly histories down to $z=4$ ({\tt CosmoRun}; Section \ref{sec:results}). 
With numerical accuracy that resolves the internal structure of a target halo ($\lesssim 100$ physical pc at $z=4$), we find that the codes overall agree well with one another in many aspects.  
We argue that, if adequately tested in accordance with our proposed calibration steps and common parameters, results of a modern high-resolution cosmological zoom-in simulations are robust and their predictive power can be maximized.
While this calibration does lead to substantial agreement on critical parameters, differences still remain between the codes --- in the properties of the gas, stars and the CGM --- due to different stellar feedback strategies adopted in each of the participating codes, as well as the diversity in implementations of the hydrodynamics.
We show that the gas distribution in the density-temperature space is globally affected by differences in the stellar feedback, particularly in the coldest and hottest gas, while achieving solid convergence in the cool and warm gas. 
We also confirm that the spatial distribution of gas metallicity from metals released in the supernova explosion is a key parameter when testing stellar feedback prescriptions in cosmological models.
This is because they play an important role in the gas cooling rates, amplifying the differences in the feedback prescriptions. 
A similar effect is observed when analyzing stellar metallicities. 
We also confirm that the expected timing discrepancies in halo mergers need to be accounted for when making code-to-code comparisons, since variations in the host's post-merger relaxation highly impacts the gas properties.
The analysis presented in this paper, that includes only five redshift epochs (i.e., $z=8,\,7,\,6,\,5$ and $4$), serves as a first presentation of our suite of cosmological zoom-in simulations, and we are currently running them down to lower redshift and saving snapshots at finer timesteps. 

It is important to briefly note a few points about our study presented in this work:  
{\it (1)}  Our comparison in this paper across different code platforms was possible only because we have established a solid baseline through rigorous calibration steps (Section \ref{sec:calibration}).
The proposed calibration procedure has enabled us to trust that any differences can only be attributed to the chosen stellar feedback prescriptions and the (relatively minor) intrinsic variations of the codes' numerics. 
{\it (2)}  The process of running cosmological simulations through multiple calibration steps and production stages has required Herculean endeavor by many {\it AGORA} members.
It was also facilitated by close discussions between the code representatives, through 3 workshops and more than 30 telecons (for the {\tt CosmoRun} simulations alone; as of May 2021), hosted by the Collaboration. 
This type of inter-platform collaboration is somewhat novel in the field of numerical cosmology. 
{\it (3)}  Throughout this invaluable learning process, participants have used {\it AGORA} as a forum to talk to and learn from one another about other codes, and sometimes surprisingly, about their own.  
Many participants have been able to improve their codes and simulation strategies.  
The new versions of {\sc Grackle} and {\tt yt} were tested on multiple code platforms during this work, providing  useful feedback to the respective developer communities.

We pride ourselves on our contribution to the galaxy formation community, by helping to maintain the  reproducibility of galaxy formation simulations in general.    
{\it AGORA} helps to raise the predictive power of numerical experiments --- this time, in particular, of cosmological zoom-in simulations --- in building and testing the theory of structure formation in Universe, thereby benefiting researchers who rely on the robustness of simulations.  
Furthermore, we have demonstrated how the multi-platform approach like {\it AGORA} could be useful to better understand how the Universe works.  
For example, in {\it AGORA}, the thermal and kinematic states of the CGM --- in which disparities exist between contemporary numerical simulations on different code platforms --- can be easily investigated with multiple codes and increased fidelity, as showcased in Section \ref{sec:results}.     
Indeed, {\it AGORA} enables a well-controlled science case in which we test various stellar feedback prescriptions and confront simulations with the ones from other codes. 
The novel infrastructure presented in this work will provide the {\it AGORA} community (or the broader simulation community) with a tool to undertake a number of new comparison projects, including the analysis of the CGM properties in simulations with different stellar feedback, the formation of clumps at high redshift, and many others. 
It should be noted that the code groups involved in other ongoing projects in {\it AGORA} or in any upcoming new projects are not limited to the seven codes that participated in this paper. 
Our Collaboration is open to the participation of new code groups, and we encourage interested community members to test their code's compatibility on their own, by adopting the common initial conditions, the common physics package, and the proposed calibration steps, and comparing their results with the ones from the models presented by the {\it AGORA} Collaboration.

\acknowledgments

We thank all of our colleagues participating in the {\it AGORA} Project for their collaborative spirit which has allowed the {\it AGORA} Collaboration to remain strong as a platform to foster and launch multiple science-oriented comparison efforts.  
We thank Aldo Rodr\'iguez-Puebla for sharing results from the abundance matching semi-empirical models, and Volker Springel for providing the original versions of {\sc Gadget-3} to be used in the {\it AGORA} Project. 
We also thank the anonymous referee for his/her insightful comments and suggestions.
This research used resources of the National Energy Research Scientific Computing Center, a DOE Office of Science User Facility supported by the Office of Science of the U.S. Department of Energy under Contract No. DE-AC02-05CH11231. 
Santi Roca-F\`abrega acknowledges support from a Spanish postdoctoral fellowship, under grant number 2017-T2/TIC-5592. 
His work has been supported by the Madrid Government (Comunidad de Madrid-Spain) under the Multiannual Agreement with Complutense University in the line Program to Stimulate Research for Young Doctors in the context of the V PRICIT (Regional Programme of Research and Technological Innovation).
He also acknowledges financial support from the Spanish Ministry of Economy and Competitiveness (MINECO) under grant number AYA2016-75808-R, AYA2017-90589-REDT and S2018/NMT-429, and from the CAM-UCM under grant number PR65/19-22462.
Ji-hoon Kim acknowledges support by Samsung Science and Technology Foundation under Project Number SSTF-BA1802-04.  
His work was also supported by the National Institute of Supercomputing and Network/Korea Institute of Science and Technology Information with supercomputing resources including technical support, grants KSC-2018-CRE-0052 and KSC-2019-CRE-0163.
Kentaro Nagamine acknowledges the support by the MEXT/JSPS KAKENHI Grant Number JP17H01111, 19H05810 \& 20H00180, as well as the travel support from the Kavli IPMU, World Premier Research Center Initiative (WPI), where part of this work was conducted.
Alessandro Lupi acknowledges funding by the MIUR under the grant PRIN 2017-MB8AEZ.
Daniel Ceverino is a Ramon-Cajal Researcher and is supported by the Ministerio de Ciencia, Innovaci\'{o}n y Universidades (MICIU/FEDER) under research grant PGC2018-094975-C21.
H\'{e}ctor Vel\'{a}zquez acknowledges support from PAPIIT-UNAM under grant number IN101918 and also by the Centro Nacional de Supercomputo (CNS-IPICYT-CONACYT). 
{\sc Art-I} simulations were performed on the {\sc Brigit/Eolo} cluster at the Centro de Proceso de Datos, Universidad Complutense de Madrid, and on the {\sc At\'ocatl}supercomputer at the LAMOD/IAUNAM. LAMOD is a collaborative effort between the IA, ICN and IQ institutes at UNAM. 
{\sc Ramses} simulations were performed on the {\sc Miztli} supercomputer at the LANACAD, Universidad Nacional Aut\'onoma de M\'exico, within the research project LANCAD-UNAM-DGTIC-151 and on the Laboratorio Nacional de Superc\'mputo del Sureste-Conacyt.
{\sc Changa} simulations were performed on the {\sc At\'ocatl} supercomputer at the Instituto de Astronom\'ia de la UNAM, and on the Extreme Science and Engineering Discovery Environment (XSEDE) allocations TG-AST20020 and TG-MCA94P018.  
XSEDE is supported by the National Science Foundation (NSF) grant ACI-1053575.
{\sc Gadget3-Osaka} simulations and analyses were performed on the XC50 systems at the Center for Computational Astrophysics (CfCA) of the National Astronomical Observatory of Japan (NAOJ), {\sc Octopus} at the Cybermedia Center, Osaka University, and {\small Oakforest-PACS} at the University of Tokyo as part of the HPCI system Research Project (hp190050, hp200041).
The publicly available {\sc Enzo} and {\tt yt} codes used in this work are the products of collaborative efforts by many independent scientists from numerous institutions around the world.  
Their commitment to open science has helped make this work possible.   

\bibliographystyle{aasjournal}
\bibliography{refs}


\end{document}